\documentclass[prr, amsfonts, amssymb, amsmath, reprint, showkeys, nofootinbib, twoside, superscriptaddress, aps, longbibliography]{revtex4-2}
\usepackage[english]{babel}
\usepackage[utf8]{inputenc}
\usepackage[colorinlistoftodos, color=green!40, prependcaption]{todonotes}
\usepackage{amsmath}
\usepackage{siunitx}
\usepackage[pdftex, pdftitle={Article}, pdfauthor={Author}]{hyperref} % For hyperlinks in the PDF

\usepackage{amsthm}
\usepackage{mathtools}
\usepackage{physics}
\usepackage{xcolor}
\usepackage{graphicx}
\usepackage[left=23mm,right=13mm,top=35mm,columnsep=15pt]{geometry} 
\usepackage{adjustbox}
\usepackage{placeins}
\usepackage[T1]{fontenc}
\usepackage{lipsum}
\usepackage{csquotes}
\usepackage{hyperref}
\usepackage{cleveref} % smart reference

\usepackage{tikz}
\usetikzlibrary{arrows,decorations.pathmorphing}

\newcommand{\maximize}{\operatornamewithlimits{maximize}}

\newcommand{\cN}{\mathcal{N}}

\usepackage{xcolor}
\usepackage[normalem]{ulem} % For \sout
\begin{document}

\title{Optimizing frequency allocation for \\fixed-frequency superconducting quantum processors}
\author{Alexis Morvan}
\thanks{Correspondence should be addressed to \href{mailto:morvan.alexis@gmail.com}{morvan.alexis@gmail.com} and \href{mailto:jmlarson@anl.gov}{jmlarson@anl.gov}}
\affiliation{Quantum Nanoelectronics Laboratory, Department of Physics, University of California at Berkeley, Berkeley, CA 94720, USA}
\affiliation{Computational Research Division, Lawrence Berkeley National Laboratory, Berkeley, CA 94720, USA}
\author{Larry Chen}
\affiliation{Quantum Nanoelectronics Laboratory, Department of Physics, University of California at Berkeley, Berkeley, CA 94720, USA}
\affiliation{Computational Research Division, Lawrence Berkeley National Laboratory, Berkeley, CA 94720, USA}
\author{Jeffrey M. Larson}
\thanks{Correspondence should be addressed to \href{mailto:morvan.alexis@gmail.com}{morvan.alexis@gmail.com} and \href{mailto:jmlarson@anl.gov}{jmlarson@anl.gov}}
\affiliation{Mathematics and Computer Science Division, Argonne National Laboratory, Lemont, IL 60439, USA}
\author{David I. Santiago}
\affiliation{Quantum Nanoelectronics Laboratory, Department of Physics, University of California at Berkeley, Berkeley, CA 94720, USA}
\author{Irfan Siddiqi}
\affiliation{Quantum Nanoelectronics Laboratory, Department of Physics, University of California at Berkeley, Berkeley, CA 94720, USA}
\affiliation{Computational Research Division, Lawrence Berkeley National Laboratory, Berkeley, CA 94720, USA}
\affiliation{Materials Sciences Division, Lawrence Berkeley National Laboratory, Berkeley, CA 94720, USA}

\date{\today} % Leave empty to omit a date

\begin{abstract}
Fixed-frequency superconducting quantum processors are one of the most mature quantum computing architectures with high-coherence qubits and simple controls. However, high-fidelity multi-qubit gates pose tight requirements on individual qubit frequencies in these processors , and these constraints are difficult to satisfy when constructing larger processors due to the large dispersion in the fabrication of Josephson junctions. In this article, we propose a mixed-integer-programming-based optimization approach that determines qubit frequencies to maximize the fabrication yield of quantum processors. We study traditional qubit and qutrit (three-level) architectures with cross-resonance interaction processors. We compare these architectures to a differential AC-Stark shift based on entanglement gates and show that our approach greatly improves the fabrication yield and also increases the scalability of these devices. Our
approach is general and can be adapted to problems where one must avoid specific frequency collisions. 
\end{abstract}

\maketitle

% Introduction
Superconducting circuits are one of the leading platforms for quantum
information processing~\cite{Arute2019-pj, jurcevic2020}, and many of the
efforts on this platform are currently directed toward scaling up to a sufficient number
of qubits that will demonstrate a clear advantage over classical computation.
In the near term, the so-called noisy intermediate-scale quantum devices and
algorithms~\cite{preskill2018} represent the exciting prospect of gaining this
advantage with a moderate number of qubits before
fault-tolerant~\cite{campbell2017} devices can be realized. One of the primary
challenges that must be addressed when scaling up such devices is how to
keep both coherence and high-fidelity control over larger numbers of qubits
without significantly increasing the complexity of these devices.

Among the different competing architectures with superconducting circuits,
fixed-frequency lattices with all-microwave control for both single- and
multiqubit operations~\cite{leek2009, chow2012,
poletto2012,chow2013,cross2015,egger2019, krinner2020} offer promising
advantages: the absence of flux tuning significantly reduces the amount of
control wiring required and allows the high coherence properties of
fixed-frequency transmons to be preserved. Recent progress on such platforms
has also shown multiqubit entanglement to realize a Toffoli-like
gate~\cite{kim2021} and uses of higher energy levels~\cite{blok2021,
morvan2021}. Despite these advantages, however, fixed-frequency architectures
are often limited to slower entangling gates and tighter frequency constraints
on the device for maintaining high-fidelity control.

The cross-resonance (CR) gate -- currently the most popular all-microwave
entangling gate -- requires that the involved transmon qubits are not detuned
by more than their anharmonicity in order to generate fast entangling
gates~\cite{rigetti2010}. On the other hand, the addressability of individual
transmons requires sufficiently large detunings between neighboring qubits to
avoid the detrimental effects of crosstalk. These constraints, combined with
the relatively large fabrication-limited frequency dispersion of the transmon,
make it extremely difficult to determine reasonable frequencies for more than a handful of qubits. 

The frequency of
a transmon qubit is determined by the critical current of the Josephson
junction (JJ) and the total capacitance of the junction. In the transmon regime ($E_J \gg E_C$), this is typically dominated by an external shunting capacitance that can generally be realized with higher precision than the critical current of the JJ. The
complexity of JJ fabrication leads to a large relative dispersion of the
critical current that carries over to the frequency of the corresponding
transmon. The best demonstrated dispersion is $\sigma_f/f = 1 \%
$~\cite{kreikebaum2020improving} without postprocessing. Recently, a
postprocessing annealing step was demonstrated that allowed for a further
reduction in this dispersion to $\sigma_f/f = 0.25 \%
$~\cite{hertzberg2020laserannealing}. Even with this improvement, however,
scaling to devices with more than a few hundred qubits will prove difficult
because of the low fabrication yield of frequency-optimal devices and will
require compromises with the standard square-lattice structure proposed for the
surface code \cite{fowler2012}.

In this article, we provide an optimization approach for maximizing the yield of
usable quantum processors for fixed-frequency architectures. We then analyze
the yield for small systems (an 8-qubit ring) and larger lattices. We
also extend our analysis to qutrit systems \cite{blok2021, morvan2021}, as well
as qubit systems using an off-resonantly driven Control-Z (CZ) gate
\cite{mitchell2021hardwareefficient} in place of the typical CR gate for
entanglement. With the extra degree of freedom given by the drive frequency of the CZ gate, we
show that one can increase the yield of these devices and scale them up to more
than 1,000 transmons without sacrificing fabrication yield.

\medskip
% Frequency collision: definition of the problem
\section{Frequency collisions}
    \begin{table*}[]
    \caption{\label{tab:constraints} Constraints imposed by the
    architecture on the frequency allocation on the graph. We group these
    constraints into three categories (described in the main text). The index
    refers to the participants and is defined by the graph underlying the
    processors. $\vec{E}$ represents the oriented graph edges, and $E$ represents the
    unoriented edges, meaning that each edge is taken into account twice in
    each direction. The spectator constraints have different participants for
    the CR and CZ cases. In the CR case, the pulse is applied only 
    to the control qubit, whereas in the CZ case the pulse is applied to both
    entangled qubits. The variable $f_d$ represents the frequency of the
    applied drive microwave. For the qutrit case we need to be able to drive
    both the $\ket{0} \rightarrow \ket{1}$ and the $\ket{1} \rightarrow
    \ket{2}$ transitions. For CZ, the frequency of the drive
    can vary. For some architectures, some constraints are
    redundant; however, they are necessary for other architectures or qutrit
    drives. The bounds values are taken from~\cite{hertzberg2020laserannealing}.}
\begin{tabular}{|ll|lcll|}
\hline
               &    &  & Definition                                                       & Participants        & Bounds \\ \hline
Addressability & A1 &  & $\left| f_i - f_j \right| \ge \delta_{A1}$                       & $(i,j) \in E$       & 17 MHz \\
               & A2 &  & $\left| f_i - f_j - \alpha_j \right| \ge \delta_{A2}$            & $(i,j) \in E$       & 30 MHz \\ \hline
Entanglement   & C1 &  & $f_{i} + \alpha_i \le f_d \le f_i $                              & $(i,j) \in \vec{E}$ & ---    \\
               & E1 &  & $\left| f_d - f_i \right| \ge \delta_{E1}$                       & $(i,j) \in \vec{E}$ & 17 MHz \\
               & E2 &  & $\left| f_d - f_i - \alpha_i \right| \ge \delta_{E2}$            & $(i,j) \in \vec{E}$ & 30 MHz \\
               & D1 &  & $\left| f_d - f_i - \alpha_i/2\right| \ge \delta_{D1}$           & $(i,j) \in \vec{E}$ & 2 MHz  \\ \hline
Spectators     & S1 &  & $\left| f_d - f_k \right| \ge \delta_{S1}$                       & $(i,j, k) \in N$    & 17 MHz \\
               & S2 &  & $\left| f_d - f_k - \alpha_k \right| \ge \delta_{S2}$            & $(i,j, k) \in N$    & 25 MHz \\
               & T1 &  & $\left| f_{d} + f_{k} - 2f_{i}-\alpha_i\right| \ge \delta_{T1} $ & $(i,j, k) \in N$    & 17 MHz \\ \hline
\end{tabular}
\end{table*}

Frequency collisions occur when an unwanted degeneracy leads to a degradation
of control fidelity for one of the native gates in a given architecture. A
simple example is when two adjacent qubits have the same $\ket{0} \mapsto
\ket{1}$ transition frequency: any unwanted couplings and/or crosstalk within
the device will lead to unwanted driving of a neighboring qubit when one is
driven, leading to a reduction in the fidelity of these operations on the
quantum processor. The presence of frequency collisions leads to a lower
fabrication yield of usable devices, which we define as the probability of
fabricating a zero-collision device given a normal distribution of frequencies
around a target frequency. To avoid such a situation, we can require that the
relevant frequencies of neighboring transmons be separated by a minimum
detuning $\delta$. This margin $\delta$ can be estimated \textit{a priori} by
simulating the gate dynamics and defining a tolerance for a target gate
fidelity requirement~\cite{magesan2020, hertzberg2020laserannealing}. Most of
the time, these frequency requirements can be described as a linear constraint
with an absolute value of the form
\begin{equation} 
  \left| f_d - f_t \right| \ge \delta ,
\end{equation}
where $f_d$ is the frequency of the drive applied and $f_t$ is the frequency
to avoid with a margin of $\delta$. \Cref{tab:constraints} lists 
these constraints divided into three categories.

\emph{Addressability} --- The first type of constraint involves the driving of
a single qubit. To ensure that applying a pulse at the $\ket{0} \mapsto
\ket{1}$ frequency of a transmon $i$ does not affect its neighbor $j$ through
unwanted crosstalk or a CR interaction, we require that the
transition frequency is different from the transition frequencies of its
nearest neighbors. These constraints are referred to as type A in
\Cref{tab:constraints}. For a qubit architecture, only the states $\ket{0}$ and
$\ket{1}$ are populated, so we only need to avoid the frequency of the
transitions that involve these states.

\emph{Entanglement constraints for CR} --- To drive a
CR entangling gate, a microwave pulse is applied to the control
transmon $i$ at the frequency of the target transition, $f_d = f_j$.
High-fidelity gates based on this effect require that the two transmons be in
the so-called straddling regime, which restricts the maximum distance between
the frequencies of adjacent transmons to be less than their anharmonicity \cite{tripathi2019, magesan2020} (see
constraint C1 in \Cref{tab:constraints}). While applying this microwave drive
to the control transmon, we require that transitions of the control transmon
be avoided (see constraints E1 and E2). Since the CR pulse is usually stronger than the single-qubit
pulses, we must also avoid the 2-photon transition (see constraint D1).

\emph{Spectators of entanglement} --- While driving an entangling gate between
two transmons, the pulse applied to one transmon can also drive the transitions
of its other neighbors $j$. These transmons are often referred to as
\textit{spectators} of the entangling gate(see constraint S1 and S2 for the 1 photons spectator collision and T1 for a 2 photons transition collision), and these classes of errors have
recently been studied in more depth \cite{sundaresan2020}.

These constraints were first defined for fixed-frequency \textit{qubit}
architectures that use the CR gate for entanglement. In the
following paragraphs, we extend these constraints to address the additional
constraints needed for \textit{qutrit} systems with the CR gate
and for \textit{qubit} systems with a differential AC-Stark shift entangling gate.

\emph{Qutrit} --- To extend these constraints to qutrit architectures, we need
to consider the addition of two drives: one for the single-qutrit gate with a
drive frequency $f_d$ at the $\ket{1} \mapsto \ket{2}$ transition and one for an
entanglement drive frequency at the $\ket{1} \mapsto \ket{2}$ frequency of the
target transmon. In addition, we need to add constraints on the drive
frequencies such that the $\ket{2} \mapsto \ket{3}$ transition frequency for
each transmon is avoided to prevent leakage. See the Supplement for an exhaustive
list of constraints. 

\emph{Differential AC-Stark shift or SiZZle} --- A control-Z gate with simultaneous AC-Stark shift
entangling gate has recently been proposed and demonstrated for both
fluxonium~\cite{xiong2021} and transmon
architectures~\cite{mitchell2021hardwareefficient, wei2021quantum} as an
alternative to the CR gate. The constraints for this gate are similar to
those for
the CR case with the modification that the drive frequency can now vary between the frequency of the control and the target transmon. As we will show, this
adds an extra degree of freedom that increases the yield of usable chips. In
addition, since there is a drive on both transmons participating in the
entangling gate, the spectator constraints must take into account the neighbors
of both transmons and not just the control transmon. See the Supplement for the
list of additional constraints. Note that this architecture has been less studied both numerically and experimentally, so we have chosen in this article pessimistic bounds that are consistent with~\cite{mitchell2021hardwareefficient}. We expect that these bounds will likely loosen as this gate is studied more. In the rest of this article, we will abbreviate this architecture as CZ architecture.

% Another important difference is the neighbors of the entanglement pulse $N$. For this architecture, the notion of target and control is not necessary, but imposes that neighbors of both transmons are taken into account into $N$, increasing the number of constraints.

% MIP solver: optimization strategie
\section{Optimization strategy}
    When constructing a quantum processor, we first need to specify the
connectivity of the device. To do so, we define a directed graph where each
node is a transmon and each edge corresponds to a possible entangling
operation. The orientation of the edge is important for the CR
gate because this specifies the directionality of the gate. Then, we need to find a
set of frequencies on these nodes that must satisfy the constraints given in
\Cref{tab:constraints}. Determining whether this type of system is feasible can
be done with modern optimization tools. Since the set of parameters giving a
feasible solution is disjoint, mixed-integer programming (MIP) is needed in this
case. We use the Gurobi solver~\cite{gurobi}
with the Python package Pyomo~\cite{pyomo}.

Initially, we found that a naive objective function yielded solutions that are on the edges of the collision-free regions, meaning that small
perturbations in frequencies often resulted in considerable collisions. Ultimately, such a naive
optimization yields a design that is not robust against the dispersion of
frequency due to inexactness in the the fabrication. To circumvent this, we have developed a
three-step approach that attempts to be a proxy for the yield: the yield should be
proportional to the distance between the target frequency and the collision
regions. Thus, we seek to maximize the distance between the target frequencies
to these avoided regions. 

To move the solution away from the border of the collision region, we introduce
a new variable per set of constraints that corresponds to the distance of the left-hand
side of the constraint definition to the threshold $\delta$ of the constraints
in \Cref{tab:constraints}. We then add a constraint that forces all of these
distances to be equal for each constraint type and each edge. This 
finds the largest hypersphere of radius $R$ that can fit inside the
zero-collision region. In a second step, we relax the requirement that all the
thresholds of each type (or rows in \Cref{tab:constraints}) be equal, but
we still require that they be equal for each edge. We also add a constraint
specifying that all of these distances are larger than the radius $R$
calculated in the previous step. This amounts to allowing for a hyperellipse
with radius $R_i \leq R$ per constraint type. In the last step, we
allow each edge to differ. See the Supplement for a more in-depth description
of the objective functions.

With a differential AC-Stark shift architecture, we must also
optimize the entanglement drive frequency since this is now a continuous
variable. To do so, we add an additional variable on each edge corresponding to
the drive frequencies for the corresponding pair of transmons.

% Small system analysis
\section{Small-scale systems}
    \begin{figure*}
    \begin{center}
        \includegraphics{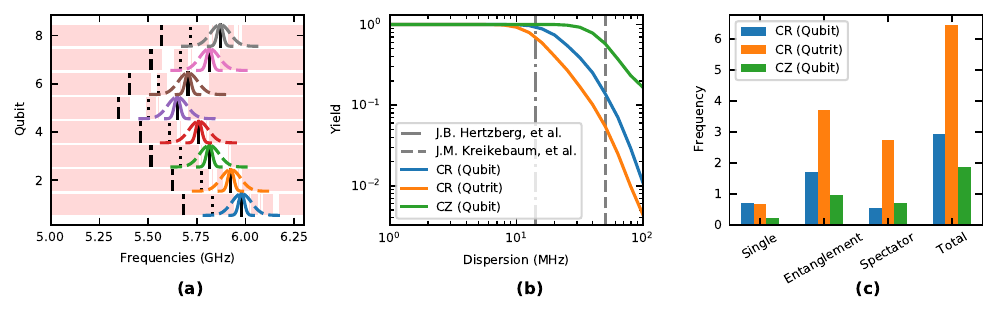}. % todo, maybe remove the with?
    \end{center}
    \caption{\label{fig:ring_yield} (a) Collision regions of an 8-qubit ring.
    The solid black line indicates the $\ket{0} \rightarrow \ket{1}$, the dashed
    black line the $\ket{1} \rightarrow \ket{2}$ transition, and the dotted line
    the 2-photon transitions $\ket{0} \rightarrow \ket{2}$. The red regions
    indicate regions considered as collisions for the device. We have plotted
    the normal distribution of frequency for the state-of-the-art dispersion
    from \cite{kreikebaum2020improving} (solid) and \cite{hertzberg2020laserannealing} (dashed).
    (b) The yield of zero-collision devices for different architectures as a function of
    average frequency dispersion.
    (c) Frequency of each category of collision for the different
    architectures on the 8-qubit ring. See main text for a discussion.}
\end{figure*}    

To build intuition about the result of the frequency optimization, we
discuss in this section the optimization of an 8-transmon ring. Such a linear
topology is the minimum connectivity possible for a quantum processor and sets
an upper limit on the fabrication yield one can expect. In
\Cref{fig:ring_yield} we present the frequency pattern obtained with the
optimization described in the preceding section, along with the allowed regions for
the frequencies of each transmon. We also present the yield as a function of
frequency dispersion, with the state-of-the-art fabrication dispersions with
and without postprocessing denoted to give a sense of achievable
yields~\cite{kreikebaum2020improving, hertzberg2020laserannealing}. 

With a qubit CR-based architecture, even a linear
topology with state-of-the-art dispersion of 50 MHz leads to a low yield
of 10\%. This yield can be increased with a postprocessing
fabrication dispersion of 14 MHz, which highlights the need for techniques such as 
laser-annealing for further reducing the frequency dispersion. In
\Cref{fig:ring_yield}c we can see the distribution of errors among the
different types of collisions. The majority of the collisions come from the
entanglement constraints; these significantly reduce the overall yield.

In \Cref{fig:ring_yield} we also present the yield for a qutrit CR-based
architecture, which has a lower yield than the corresponding qubit architecture
due to the increased number of constraints. Here we optimize
the frequency layout for the qubit constraints and simply
calculate the yield on the qutrit architecture with the additional constraints
added. Note that the frequency of single-transmon collisions
remains the same for both the qubit and qutrit architectures because the optimization fixes the same anharmonicity $\alpha$ for every
transmon and the energy spectrum of the transmon. This suggests that while
fabricating collision-free devices using qutrit entanglement is
difficult, single-qutrit gates can be implemented with few frequency collisions
on a qubit-optimized device.

Finally, we also present the yield for a qubit architecture based on the
differential AC-Stark shift gate. Because of the extra degree of freedom offered by
the frequency of the entangling drive, the yield is much higher, and we see a
larger plateau region where the yield is approximately perfect. To calculate
the yield of this specific architecture, we sampled random frequencies around
the target layout frequency, and for each random sample, we used the optimizer to
find a new set of optimal driving frequencies for the entangling gate. For this
architecture, state-of-the-art fabrication without postprocessing is already
capable of reaching a high yield of 90\%, and almost perfect yield can be
attained with additional postprocessing. Compared with the CR-based
architecture, the collision rates decrease for both single- and two-qubit gates
because of the relaxation of the drive frequency constraints for entangling
operations.

% lattice analysis
\section{Scaling to lattices}
    \begin{figure*}
    \begin{center}
        \includegraphics[width= \textwidth]{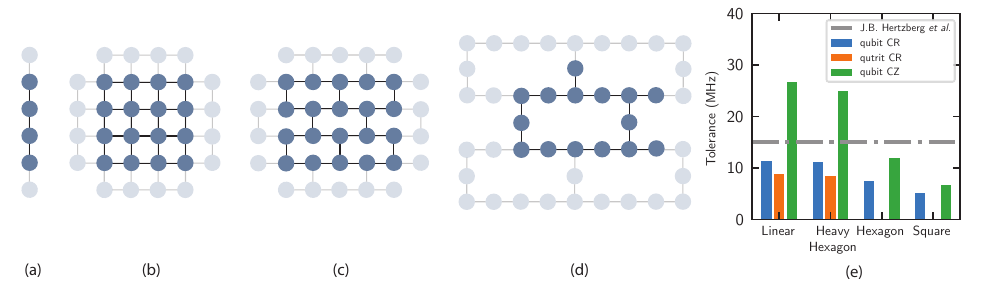}. % todo, maybe remove the with?
    \end{center}
    \caption{\label{fig:lattice} (a-d) Common lattice layouts. The link represents the connectivity or the possibility of realizing entanglement between its two qubits. In practice, we simulate the behavior of a unit cell with periodic boundary conditions to extract the yield. (a) A linear chain with a unit cell of 8 sites. (b) A square lattice with a unit cell of 16 sites. (c) A hexagonal lattice of 20 sites. (d) A heavy hexagon lattice with a unit cell of 15 sites. (e)
    Tolerance on the dispersion of the frequency to achieve a 10\% yield on
    different lattice topologies. For the qutrit case on the hexagon and square
    lattices, our qubit solution was not compatible with qutrit constraints.
    The dashed line indicates the dispersion obtained by
    \cite{hertzberg2020laserannealing} with laser-annealing.}
\end{figure*}    

As the number of transmons on a processor increases, so does the number of
collisions. For large systems, individually optimizing all the frequencies on a
chip becomes intractable simply because of the number of variables and
constraints. Instead, a smaller unit cell of frequencies must be optimized and
then tiled to generate a larger lattice. To construct such a solution, we start
by defining the unit cell to be tiled. To ensure that the tiling is possible, we
optimize the frequency layout with periodic boundary conditions, ensuring that
the solution is also valid for the larger lattice. Since calculating the yield is
quick for the CR case, we can directly calculate the zero-collision yield for
these large lattices. \Cref{fig:lattice} shows the yield for different lattices
with different connectivities: a square lattice for the surface code, a heavy
hexagonal lattice used by IBM \cite{chamberland2020}, and a hexagonal lattice
for comparison with its heavy counterpart. 

% We also present the tolerance for the same chip for qutrit. We note that for
% the higher connectivity lattice, we were not able to find a solution for
% qutrit. The spectator error for qutrit S3 seems to be incompatible. 

When scaling to larger lattices from a given unit cell, the yield of the system
can be estimated up to boundary effects with a simple scaling law,
\begin{equation}
    y = y_{\text{m}}^{N/n_{\text{m}}} = y_0^N,
\end{equation}
where $y$ is the yield of the larger system, $y_m$ is the yield of the unit cell
with periodic boundary conditions, $N$ is the total number of transmons in the
system and $n_m$ is the number of transmons in the unit cell. Since the
boundary effects will only reduce the number of potential collisions, this
provides a lower bound on the yield, even for smaller system sizes. This method
provides a way for comparing the scalability of different types of lattices.

In \Cref{fig:lattice} we show the frequency dispersion required to
achieve a yield of $10\%$ for a 1,000-qubit lattice of each type.
Unsurprisingly, the more densely connected lattice has a lower yield.
We note that given the thresholds listed in \Cref{tab:constraints} for a
CR-based qubit architecture, none of these lattices can be fabricated with such
a yield given even the best-demonstrated frequency dispersion with
postprocessing. This shows that a collision-free 1,000-qubit device is likely
unachievable for the CR-based architecture without a significant improvement in
fabrication precision or a relaxation of the constraints. In contrast, the
AC-Stark shift-based qubit architecture can scale to 1,000 qubits for two of the four
lattices with a frequency dispersion already attainable with postprocessing.

% lattice analysis
\section{Conclusion}
    In this article, we have discussed the optimization of the frequency layout of fixed-frequency superconducting quantum processors. We have discussed the yield of the Cross-Resonance architecture for qubit and qutrit. We have also shown that using the differential AC-Stark shift to realize a CZ entanglement gate gives an extra degree of freedom on the entanglement drive frequency and improves the yield of this architecture, thus improving the scaling possibility of this architecture. 
Fixed-frequency quantum processor with transmon is currently one of the most developed architectures for multi-qubit systems, however, other qubit designs like fluxonium \cite{xiong2021} can also realize quantum processors and will have different frequency constraints that can potentially give an advantage over transmon in terms of yield.

\vspace{1em}
\section*{Acknowledgment}
    We gratefully acknowledge the conversations and insights of R. Naik and B. Mitchell. We also thank G. Koolstra and L. Nguyen for their comment on the manuscript.
    We thank three anonymous referees for their constructive feedback that improved an early version of this manuscript.
    This work was supported by the Quantum Testbed Program of the Advanced
    Scientific Computing Research program for Basic Energy Sciences and the Office of
    Science of the U.S. Department of Energy under Contract Nos.~DE-AC02-05CH11231 and DE-AC02-06CH11357.
    This work was supported by the U.S. Department of Energy, Office of Science,
    Office of Advanced Scientific Computing Research, Accelerated Research for
    Quantum Computing program.

\bibliographystyle{apsrev4-2}
\bibliography{biblio}%

\begin{thebibliography}{27}%
\makeatletter
\providecommand \@ifxundefined [1]{%
 \@ifx{#1\undefined}
}%
\providecommand \@ifnum [1]{%
 \ifnum #1\expandafter \@firstoftwo
 \else \expandafter \@secondoftwo
 \fi
}%
\providecommand \@ifx [1]{%
 \ifx #1\expandafter \@firstoftwo
 \else \expandafter \@secondoftwo
 \fi
}%
\providecommand \natexlab [1]{#1}%
\providecommand \enquote  [1]{``#1''}%
\providecommand \bibnamefont  [1]{#1}%
\providecommand \bibfnamefont [1]{#1}%
\providecommand \citenamefont [1]{#1}%
\providecommand \href@noop [0]{\@secondoftwo}%
\providecommand \href [0]{\begingroup \@sanitize@url \@href}%
\providecommand \@href[1]{\@@startlink{#1}\@@href}%
\providecommand \@@href[1]{\endgroup#1\@@endlink}%
\providecommand \@sanitize@url [0]{\catcode `\\12\catcode `\$12\catcode
  `\&12\catcode `\#12\catcode `\^12\catcode `\_12\catcode `\%12\relax}%
\providecommand \@@startlink[1]{}%
\providecommand \@@endlink[0]{}%
\providecommand \url  [0]{\begingroup\@sanitize@url \@url }%
\providecommand \@url [1]{\endgroup\@href {#1}{\urlprefix }}%
\providecommand \urlprefix  [0]{URL }%
\providecommand \Eprint [0]{\href }%
\providecommand \doibase [0]{https://doi.org/}%
\providecommand \selectlanguage [0]{\@gobble}%
\providecommand \bibinfo  [0]{\@secondoftwo}%
\providecommand \bibfield  [0]{\@secondoftwo}%
\providecommand \translation [1]{[#1]}%
\providecommand \BibitemOpen [0]{}%
\providecommand \bibitemStop [0]{}%
\providecommand \bibitemNoStop [0]{.\EOS\space}%
\providecommand \EOS [0]{\spacefactor3000\relax}%
\providecommand \BibitemShut  [1]{\csname bibitem#1\endcsname}%
\let\auto@bib@innerbib\@empty
%</preamble>
\bibitem [{\citenamefont {Arute}\ \emph {et~al.}(2019)\citenamefont {Arute},
  \citenamefont {Arya}, \citenamefont {Babbush}, \citenamefont {Bacon},
  \citenamefont {Bardin}, \citenamefont {Barends}, \citenamefont {Biswas},
  \citenamefont {Boixo}, \citenamefont {Brandao}, \citenamefont {Buell},
  \citenamefont {Burkett}, \citenamefont {Chen}, \citenamefont {Chen},
  \citenamefont {Chiaro}, \citenamefont {Collins}, \citenamefont {Courtney},
  \citenamefont {Dunsworth}, \citenamefont {Farhi}, \citenamefont {Foxen},
  \citenamefont {Fowler}, \citenamefont {Gidney}, \citenamefont {Giustina},
  \citenamefont {Graff}, \citenamefont {Guerin}, \citenamefont {Habegger},
  \citenamefont {Harrigan}, \citenamefont {Hartmann}, \citenamefont {Ho},
  \citenamefont {Hoffmann}, \citenamefont {Huang}, \citenamefont {Humble},
  \citenamefont {Isakov}, \citenamefont {Jeffrey}, \citenamefont {Jiang},
  \citenamefont {Kafri}, \citenamefont {Kechedzhi}, \citenamefont {Kelly},
  \citenamefont {Klimov}, \citenamefont {Knysh}, \citenamefont {Korotkov},
  \citenamefont {Kostritsa}, \citenamefont {Landhuis}, \citenamefont
  {Lindmark}, \citenamefont {Lucero}, \citenamefont {Lyakh}, \citenamefont
  {Mandr{\`a}}, \citenamefont {McClean}, \citenamefont {McEwen}, \citenamefont
  {Megrant}, \citenamefont {Mi}, \citenamefont {Michielsen}, \citenamefont
  {Mohseni}, \citenamefont {Mutus}, \citenamefont {Naaman}, \citenamefont
  {Neeley}, \citenamefont {Neill}, \citenamefont {Niu}, \citenamefont {Ostby},
  \citenamefont {Petukhov}, \citenamefont {Platt}, \citenamefont {Quintana},
  \citenamefont {Rieffel}, \citenamefont {Roushan}, \citenamefont {Rubin},
  \citenamefont {Sank}, \citenamefont {Satzinger}, \citenamefont {Smelyanskiy},
  \citenamefont {Sung}, \citenamefont {Trevithick}, \citenamefont
  {Vainsencher}, \citenamefont {Villalonga}, \citenamefont {White},
  \citenamefont {Yao}, \citenamefont {Yeh}, \citenamefont {Zalcman},
  \citenamefont {Neven},\ and\ \citenamefont {Martinis}}]{Arute2019-pj}%
  \BibitemOpen
  \bibfield  {author} {\bibinfo {author} {\bibfnamefont {F.}~\bibnamefont
  {Arute}}, \bibinfo {author} {\bibfnamefont {K.}~\bibnamefont {Arya}},
  \bibinfo {author} {\bibfnamefont {R.}~\bibnamefont {Babbush}}, \bibinfo
  {author} {\bibfnamefont {D.}~\bibnamefont {Bacon}}, \bibinfo {author}
  {\bibfnamefont {J.~C.}\ \bibnamefont {Bardin}}, \bibinfo {author}
  {\bibfnamefont {R.}~\bibnamefont {Barends}}, \bibinfo {author} {\bibfnamefont
  {R.}~\bibnamefont {Biswas}}, \bibinfo {author} {\bibfnamefont
  {S.}~\bibnamefont {Boixo}}, \bibinfo {author} {\bibfnamefont {F.~G. S.~L.}\
  \bibnamefont {Brandao}}, \bibinfo {author} {\bibfnamefont {D.~A.}\
  \bibnamefont {Buell}}, \bibinfo {author} {\bibfnamefont {B.}~\bibnamefont
  {Burkett}}, \bibinfo {author} {\bibfnamefont {Y.}~\bibnamefont {Chen}},
  \bibinfo {author} {\bibfnamefont {Z.}~\bibnamefont {Chen}}, \bibinfo {author}
  {\bibfnamefont {B.}~\bibnamefont {Chiaro}}, \bibinfo {author} {\bibfnamefont
  {R.}~\bibnamefont {Collins}}, \bibinfo {author} {\bibfnamefont
  {W.}~\bibnamefont {Courtney}}, \bibinfo {author} {\bibfnamefont
  {A.}~\bibnamefont {Dunsworth}}, \bibinfo {author} {\bibfnamefont
  {E.}~\bibnamefont {Farhi}}, \bibinfo {author} {\bibfnamefont
  {B.}~\bibnamefont {Foxen}}, \bibinfo {author} {\bibfnamefont
  {A.}~\bibnamefont {Fowler}}, \bibinfo {author} {\bibfnamefont
  {C.}~\bibnamefont {Gidney}}, \bibinfo {author} {\bibfnamefont
  {M.}~\bibnamefont {Giustina}}, \bibinfo {author} {\bibfnamefont
  {R.}~\bibnamefont {Graff}}, \bibinfo {author} {\bibfnamefont
  {K.}~\bibnamefont {Guerin}}, \bibinfo {author} {\bibfnamefont
  {S.}~\bibnamefont {Habegger}}, \bibinfo {author} {\bibfnamefont {M.~P.}\
  \bibnamefont {Harrigan}}, \bibinfo {author} {\bibfnamefont {M.~J.}\
  \bibnamefont {Hartmann}}, \bibinfo {author} {\bibfnamefont {A.}~\bibnamefont
  {Ho}}, \bibinfo {author} {\bibfnamefont {M.}~\bibnamefont {Hoffmann}},
  \bibinfo {author} {\bibfnamefont {T.}~\bibnamefont {Huang}}, \bibinfo
  {author} {\bibfnamefont {T.~S.}\ \bibnamefont {Humble}}, \bibinfo {author}
  {\bibfnamefont {S.~V.}\ \bibnamefont {Isakov}}, \bibinfo {author}
  {\bibfnamefont {E.}~\bibnamefont {Jeffrey}}, \bibinfo {author} {\bibfnamefont
  {Z.}~\bibnamefont {Jiang}}, \bibinfo {author} {\bibfnamefont
  {D.}~\bibnamefont {Kafri}}, \bibinfo {author} {\bibfnamefont
  {K.}~\bibnamefont {Kechedzhi}}, \bibinfo {author} {\bibfnamefont
  {J.}~\bibnamefont {Kelly}}, \bibinfo {author} {\bibfnamefont {P.~V.}\
  \bibnamefont {Klimov}}, \bibinfo {author} {\bibfnamefont {S.}~\bibnamefont
  {Knysh}}, \bibinfo {author} {\bibfnamefont {A.}~\bibnamefont {Korotkov}},
  \bibinfo {author} {\bibfnamefont {F.}~\bibnamefont {Kostritsa}}, \bibinfo
  {author} {\bibfnamefont {D.}~\bibnamefont {Landhuis}}, \bibinfo {author}
  {\bibfnamefont {M.}~\bibnamefont {Lindmark}}, \bibinfo {author}
  {\bibfnamefont {E.}~\bibnamefont {Lucero}}, \bibinfo {author} {\bibfnamefont
  {D.}~\bibnamefont {Lyakh}}, \bibinfo {author} {\bibfnamefont
  {S.}~\bibnamefont {Mandr{\`a}}}, \bibinfo {author} {\bibfnamefont {J.~R.}\
  \bibnamefont {McClean}}, \bibinfo {author} {\bibfnamefont {M.}~\bibnamefont
  {McEwen}}, \bibinfo {author} {\bibfnamefont {A.}~\bibnamefont {Megrant}},
  \bibinfo {author} {\bibfnamefont {X.}~\bibnamefont {Mi}}, \bibinfo {author}
  {\bibfnamefont {K.}~\bibnamefont {Michielsen}}, \bibinfo {author}
  {\bibfnamefont {M.}~\bibnamefont {Mohseni}}, \bibinfo {author} {\bibfnamefont
  {J.}~\bibnamefont {Mutus}}, \bibinfo {author} {\bibfnamefont
  {O.}~\bibnamefont {Naaman}}, \bibinfo {author} {\bibfnamefont
  {M.}~\bibnamefont {Neeley}}, \bibinfo {author} {\bibfnamefont
  {C.}~\bibnamefont {Neill}}, \bibinfo {author} {\bibfnamefont {M.~Y.}\
  \bibnamefont {Niu}}, \bibinfo {author} {\bibfnamefont {E.}~\bibnamefont
  {Ostby}}, \bibinfo {author} {\bibfnamefont {A.}~\bibnamefont {Petukhov}},
  \bibinfo {author} {\bibfnamefont {J.~C.}\ \bibnamefont {Platt}}, \bibinfo
  {author} {\bibfnamefont {C.}~\bibnamefont {Quintana}}, \bibinfo {author}
  {\bibfnamefont {E.~G.}\ \bibnamefont {Rieffel}}, \bibinfo {author}
  {\bibfnamefont {P.}~\bibnamefont {Roushan}}, \bibinfo {author} {\bibfnamefont
  {N.~C.}\ \bibnamefont {Rubin}}, \bibinfo {author} {\bibfnamefont
  {D.}~\bibnamefont {Sank}}, \bibinfo {author} {\bibfnamefont {K.~J.}\
  \bibnamefont {Satzinger}}, \bibinfo {author} {\bibfnamefont {V.}~\bibnamefont
  {Smelyanskiy}}, \bibinfo {author} {\bibfnamefont {K.~J.}\ \bibnamefont
  {Sung}}, \bibinfo {author} {\bibfnamefont {M.~D.}\ \bibnamefont
  {Trevithick}}, \bibinfo {author} {\bibfnamefont {A.}~\bibnamefont
  {Vainsencher}}, \bibinfo {author} {\bibfnamefont {B.}~\bibnamefont
  {Villalonga}}, \bibinfo {author} {\bibfnamefont {T.}~\bibnamefont {White}},
  \bibinfo {author} {\bibfnamefont {Z.~J.}\ \bibnamefont {Yao}}, \bibinfo
  {author} {\bibfnamefont {P.}~\bibnamefont {Yeh}}, \bibinfo {author}
  {\bibfnamefont {A.}~\bibnamefont {Zalcman}}, \bibinfo {author} {\bibfnamefont
  {H.}~\bibnamefont {Neven}},\ and\ \bibinfo {author} {\bibfnamefont {J.~M.}\
  \bibnamefont {Martinis}},\ }\href {https://doi.org/10.1038/s41586-019-1666-5}
  {\bibfield  {journal} {\bibinfo  {journal} {Nature}\ }\textbf {\bibinfo
  {volume} {574}},\ \bibinfo {pages} {505} (\bibinfo {year}
  {2019})}\BibitemShut {NoStop}%
\bibitem [{\citenamefont {Jurcevic}\ \emph {et~al.}(2021)\citenamefont
  {Jurcevic}, \citenamefont {Javadi-Abhari}, \citenamefont {Bishop},
  \citenamefont {Lauer}, \citenamefont {Bogorin}, \citenamefont {Brink},
  \citenamefont {Capelluto}, \citenamefont {G{\"u}nl{\"u}k}, \citenamefont
  {Itoko}, \citenamefont {Kanazawa} \emph {et~al.}}]{jurcevic2020}%
  \BibitemOpen
  \bibfield  {author} {\bibinfo {author} {\bibfnamefont {P.}~\bibnamefont
  {Jurcevic}}, \bibinfo {author} {\bibfnamefont {A.}~\bibnamefont
  {Javadi-Abhari}}, \bibinfo {author} {\bibfnamefont {L.~S.}\ \bibnamefont
  {Bishop}}, \bibinfo {author} {\bibfnamefont {I.}~\bibnamefont {Lauer}},
  \bibinfo {author} {\bibfnamefont {D.~F.}\ \bibnamefont {Bogorin}}, \bibinfo
  {author} {\bibfnamefont {M.}~\bibnamefont {Brink}}, \bibinfo {author}
  {\bibfnamefont {L.}~\bibnamefont {Capelluto}}, \bibinfo {author}
  {\bibfnamefont {O.}~\bibnamefont {G{\"u}nl{\"u}k}}, \bibinfo {author}
  {\bibfnamefont {T.}~\bibnamefont {Itoko}}, \bibinfo {author} {\bibfnamefont
  {N.}~\bibnamefont {Kanazawa}}, \emph {et~al.},\ }\href
  {https://doi.org/10.1088/2058-9565/abe519} {\bibfield  {journal} {\bibinfo
  {journal} {Quantum Science and Technology}\ }\textbf {\bibinfo {volume}
  {6}},\ \bibinfo {pages} {025020} (\bibinfo {year} {2021})}\BibitemShut
  {NoStop}%
\bibitem [{\citenamefont {Preskill}(2018)}]{preskill2018}%
  \BibitemOpen
  \bibfield  {author} {\bibinfo {author} {\bibfnamefont {J.}~\bibnamefont
  {Preskill}},\ }\href {https://doi.org/10.22331/q-2018-08-06-79} {\bibfield
  {journal} {\bibinfo  {journal} {Quantum}\ }\textbf {\bibinfo {volume} {2}},\
  \bibinfo {pages} {79} (\bibinfo {year} {2018})}\BibitemShut {NoStop}%
\bibitem [{\citenamefont {Campbell}\ \emph {et~al.}(2017)\citenamefont
  {Campbell}, \citenamefont {Terhal},\ and\ \citenamefont
  {Vuillot}}]{campbell2017}%
  \BibitemOpen
  \bibfield  {author} {\bibinfo {author} {\bibfnamefont {E.~T.}\ \bibnamefont
  {Campbell}}, \bibinfo {author} {\bibfnamefont {B.~M.}\ \bibnamefont
  {Terhal}},\ and\ \bibinfo {author} {\bibfnamefont {C.}~\bibnamefont
  {Vuillot}},\ }\href {https://doi.org/10.1038/nature23460} {\bibfield
  {journal} {\bibinfo  {journal} {Nature}\ }\textbf {\bibinfo {volume} {549}},\
  \bibinfo {pages} {172–179} (\bibinfo {year} {2017})}\BibitemShut {NoStop}%
\bibitem [{\citenamefont {Leek}\ \emph {et~al.}(2009)\citenamefont {Leek},
  \citenamefont {Filipp}, \citenamefont {Maurer}, \citenamefont {Baur},
  \citenamefont {Bianchetti}, \citenamefont {Fink}, \citenamefont {G\"oppl},
  \citenamefont {Steffen},\ and\ \citenamefont {Wallraff}}]{leek2009}%
  \BibitemOpen
  \bibfield  {author} {\bibinfo {author} {\bibfnamefont {P.~J.}\ \bibnamefont
  {Leek}}, \bibinfo {author} {\bibfnamefont {S.}~\bibnamefont {Filipp}},
  \bibinfo {author} {\bibfnamefont {P.}~\bibnamefont {Maurer}}, \bibinfo
  {author} {\bibfnamefont {M.}~\bibnamefont {Baur}}, \bibinfo {author}
  {\bibfnamefont {R.}~\bibnamefont {Bianchetti}}, \bibinfo {author}
  {\bibfnamefont {J.~M.}\ \bibnamefont {Fink}}, \bibinfo {author}
  {\bibfnamefont {M.}~\bibnamefont {G\"oppl}}, \bibinfo {author} {\bibfnamefont
  {L.}~\bibnamefont {Steffen}},\ and\ \bibinfo {author} {\bibfnamefont
  {A.}~\bibnamefont {Wallraff}},\ }\href
  {https://doi.org/10.1103/PhysRevB.79.180511} {\bibfield  {journal} {\bibinfo
  {journal} {Phys. Rev. B}\ }\textbf {\bibinfo {volume} {79}},\ \bibinfo
  {pages} {180511} (\bibinfo {year} {2009})}\BibitemShut {NoStop}%
\bibitem [{\citenamefont {Chow}\ \emph {et~al.}(2012)\citenamefont {Chow},
  \citenamefont {Gambetta}, \citenamefont {C\'orcoles}, \citenamefont {Merkel},
  \citenamefont {Smolin}, \citenamefont {Rigetti}, \citenamefont {Poletto},
  \citenamefont {Keefe}, \citenamefont {Rothwell}, \citenamefont {Rozen},
  \citenamefont {Ketchen},\ and\ \citenamefont {Steffen}}]{chow2012}%
  \BibitemOpen
  \bibfield  {author} {\bibinfo {author} {\bibfnamefont {J.~M.}\ \bibnamefont
  {Chow}}, \bibinfo {author} {\bibfnamefont {J.~M.}\ \bibnamefont {Gambetta}},
  \bibinfo {author} {\bibfnamefont {A.~D.}\ \bibnamefont {C\'orcoles}},
  \bibinfo {author} {\bibfnamefont {S.~T.}\ \bibnamefont {Merkel}}, \bibinfo
  {author} {\bibfnamefont {J.~A.}\ \bibnamefont {Smolin}}, \bibinfo {author}
  {\bibfnamefont {C.}~\bibnamefont {Rigetti}}, \bibinfo {author} {\bibfnamefont
  {S.}~\bibnamefont {Poletto}}, \bibinfo {author} {\bibfnamefont {G.~A.}\
  \bibnamefont {Keefe}}, \bibinfo {author} {\bibfnamefont {M.~B.}\ \bibnamefont
  {Rothwell}}, \bibinfo {author} {\bibfnamefont {J.~R.}\ \bibnamefont {Rozen}},
  \bibinfo {author} {\bibfnamefont {M.~B.}\ \bibnamefont {Ketchen}},\ and\
  \bibinfo {author} {\bibfnamefont {M.}~\bibnamefont {Steffen}},\ }\href
  {https://doi.org/10.1103/PhysRevLett.109.060501} {\bibfield  {journal}
  {\bibinfo  {journal} {Phys. Rev. Lett.}\ }\textbf {\bibinfo {volume} {109}},\
  \bibinfo {pages} {060501} (\bibinfo {year} {2012})}\BibitemShut {NoStop}%
\bibitem [{\citenamefont {Poletto}\ \emph {et~al.}(2012)\citenamefont
  {Poletto}, \citenamefont {Gambetta}, \citenamefont {Merkel}, \citenamefont
  {Smolin}, \citenamefont {Chow}, \citenamefont {C\'orcoles}, \citenamefont
  {Keefe}, \citenamefont {Rothwell}, \citenamefont {Rozen}, \citenamefont
  {Abraham}, \citenamefont {Rigetti},\ and\ \citenamefont
  {Steffen}}]{poletto2012}%
  \BibitemOpen
  \bibfield  {author} {\bibinfo {author} {\bibfnamefont {S.}~\bibnamefont
  {Poletto}}, \bibinfo {author} {\bibfnamefont {J.~M.}\ \bibnamefont
  {Gambetta}}, \bibinfo {author} {\bibfnamefont {S.~T.}\ \bibnamefont
  {Merkel}}, \bibinfo {author} {\bibfnamefont {J.~A.}\ \bibnamefont {Smolin}},
  \bibinfo {author} {\bibfnamefont {J.~M.}\ \bibnamefont {Chow}}, \bibinfo
  {author} {\bibfnamefont {A.~D.}\ \bibnamefont {C\'orcoles}}, \bibinfo
  {author} {\bibfnamefont {G.~A.}\ \bibnamefont {Keefe}}, \bibinfo {author}
  {\bibfnamefont {M.~B.}\ \bibnamefont {Rothwell}}, \bibinfo {author}
  {\bibfnamefont {J.~R.}\ \bibnamefont {Rozen}}, \bibinfo {author}
  {\bibfnamefont {D.~W.}\ \bibnamefont {Abraham}}, \bibinfo {author}
  {\bibfnamefont {C.}~\bibnamefont {Rigetti}},\ and\ \bibinfo {author}
  {\bibfnamefont {M.}~\bibnamefont {Steffen}},\ }\href
  {https://doi.org/10.1103/PhysRevLett.109.240505} {\bibfield  {journal}
  {\bibinfo  {journal} {Phys. Rev. Lett.}\ }\textbf {\bibinfo {volume} {109}},\
  \bibinfo {pages} {240505} (\bibinfo {year} {2012})}\BibitemShut {NoStop}%
\bibitem [{\citenamefont {Chow}\ \emph {et~al.}(2013)\citenamefont {Chow},
  \citenamefont {Gambetta}, \citenamefont {Cross}, \citenamefont {Merkel},
  \citenamefont {Rigetti},\ and\ \citenamefont {Steffen}}]{chow2013}%
  \BibitemOpen
  \bibfield  {author} {\bibinfo {author} {\bibfnamefont {J.~M.}\ \bibnamefont
  {Chow}}, \bibinfo {author} {\bibfnamefont {J.~M.}\ \bibnamefont {Gambetta}},
  \bibinfo {author} {\bibfnamefont {A.~W.}\ \bibnamefont {Cross}}, \bibinfo
  {author} {\bibfnamefont {S.~T.}\ \bibnamefont {Merkel}}, \bibinfo {author}
  {\bibfnamefont {C.}~\bibnamefont {Rigetti}},\ and\ \bibinfo {author}
  {\bibfnamefont {M.}~\bibnamefont {Steffen}},\ }\href
  {https://doi.org/10.1088/1367-2630/15/11/115012} {\bibfield  {journal}
  {\bibinfo  {journal} {New J. Phys.}\ }\textbf {\bibinfo {volume} {15}},\
  \bibinfo {pages} {115012} (\bibinfo {year} {2013})}\BibitemShut {NoStop}%
\bibitem [{\citenamefont {Cross}\ and\ \citenamefont
  {Gambetta}(2015)}]{cross2015}%
  \BibitemOpen
  \bibfield  {author} {\bibinfo {author} {\bibfnamefont {A.~W.}\ \bibnamefont
  {Cross}}\ and\ \bibinfo {author} {\bibfnamefont {J.~M.}\ \bibnamefont
  {Gambetta}},\ }\href {https://doi.org/10.1103/PhysRevA.91.032325} {\bibfield
  {journal} {\bibinfo  {journal} {Phys. Rev. A}\ }\textbf {\bibinfo {volume}
  {91}},\ \bibinfo {pages} {032325} (\bibinfo {year} {2015})}\BibitemShut
  {NoStop}%
\bibitem [{\citenamefont {Egger}\ \emph {et~al.}(2019)\citenamefont {Egger},
  \citenamefont {Ganzhorn}, \citenamefont {Salis}, \citenamefont {Fuhrer},
  \citenamefont {M\"uller}, \citenamefont {Barkoutsos}, \citenamefont {Moll},
  \citenamefont {Tavernelli},\ and\ \citenamefont {Filipp}}]{egger2019}%
  \BibitemOpen
  \bibfield  {author} {\bibinfo {author} {\bibfnamefont {D.}~\bibnamefont
  {Egger}}, \bibinfo {author} {\bibfnamefont {M.}~\bibnamefont {Ganzhorn}},
  \bibinfo {author} {\bibfnamefont {G.}~\bibnamefont {Salis}}, \bibinfo
  {author} {\bibfnamefont {A.}~\bibnamefont {Fuhrer}}, \bibinfo {author}
  {\bibfnamefont {P.}~\bibnamefont {M\"uller}}, \bibinfo {author}
  {\bibfnamefont {P.}~\bibnamefont {Barkoutsos}}, \bibinfo {author}
  {\bibfnamefont {N.}~\bibnamefont {Moll}}, \bibinfo {author} {\bibfnamefont
  {I.}~\bibnamefont {Tavernelli}},\ and\ \bibinfo {author} {\bibfnamefont
  {S.}~\bibnamefont {Filipp}},\ }\href
  {https://doi.org/10.1103/PhysRevApplied.11.014017} {\bibfield  {journal}
  {\bibinfo  {journal} {Phys. Rev. Appl.}\ }\textbf {\bibinfo {volume} {11}},\
  \bibinfo {pages} {014017} (\bibinfo {year} {2019})}\BibitemShut {NoStop}%
\bibitem [{\citenamefont {Krinner}\ \emph {et~al.}(2020)\citenamefont
  {Krinner}, \citenamefont {Kurpiers}, \citenamefont {Royer}, \citenamefont
  {Magnard}, \citenamefont {Tsitsilin}, \citenamefont {Besse}, \citenamefont
  {Remm}, \citenamefont {Blais},\ and\ \citenamefont {Wallraff}}]{krinner2020}%
  \BibitemOpen
  \bibfield  {author} {\bibinfo {author} {\bibfnamefont {S.}~\bibnamefont
  {Krinner}}, \bibinfo {author} {\bibfnamefont {P.}~\bibnamefont {Kurpiers}},
  \bibinfo {author} {\bibfnamefont {B.}~\bibnamefont {Royer}}, \bibinfo
  {author} {\bibfnamefont {P.}~\bibnamefont {Magnard}}, \bibinfo {author}
  {\bibfnamefont {I.}~\bibnamefont {Tsitsilin}}, \bibinfo {author}
  {\bibfnamefont {J.-C.}\ \bibnamefont {Besse}}, \bibinfo {author}
  {\bibfnamefont {A.}~\bibnamefont {Remm}}, \bibinfo {author} {\bibfnamefont
  {A.}~\bibnamefont {Blais}},\ and\ \bibinfo {author} {\bibfnamefont
  {A.}~\bibnamefont {Wallraff}},\ }\href
  {http://dx.doi.org/10.1103/PhysRevApplied.14.044039} {\bibfield  {journal}
  {\bibinfo  {journal} {Phys. Rev. Appl.}\ }\textbf {\bibinfo {volume} {14}}
  (\bibinfo {year} {2020})}\BibitemShut {NoStop}%
\bibitem [{\citenamefont {Kim}\ \emph {et~al.}(2021)\citenamefont {Kim},
  \citenamefont {Morvan}, \citenamefont {Nguyen}, \citenamefont {Naik},
  \citenamefont {Jünger}, \citenamefont {Chen}, \citenamefont {Kreikebaum},
  \citenamefont {Santiago},\ and\ \citenamefont {Siddiqi}}]{kim2021}%
  \BibitemOpen
  \bibfield  {author} {\bibinfo {author} {\bibfnamefont {Y.}~\bibnamefont
  {Kim}}, \bibinfo {author} {\bibfnamefont {A.}~\bibnamefont {Morvan}},
  \bibinfo {author} {\bibfnamefont {L.~B.}\ \bibnamefont {Nguyen}}, \bibinfo
  {author} {\bibfnamefont {R.~K.}\ \bibnamefont {Naik}}, \bibinfo {author}
  {\bibfnamefont {C.}~\bibnamefont {Jünger}}, \bibinfo {author} {\bibfnamefont
  {L.}~\bibnamefont {Chen}}, \bibinfo {author} {\bibfnamefont {J.~M.}\
  \bibnamefont {Kreikebaum}}, \bibinfo {author} {\bibfnamefont {D.~I.}\
  \bibnamefont {Santiago}},\ and\ \bibinfo {author} {\bibfnamefont
  {I.}~\bibnamefont {Siddiqi}},\ }\href@noop {} {\bibinfo {title}
  {High-fidelity $i${Toffoli} gate for fixed-frequency superconducting qubits}}
  (\bibinfo {year} {2021}),\ \Eprint {https://arxiv.org/abs/2108.10288}
  {arXiv:2108.10288 [quant-ph]} \BibitemShut {NoStop}%
\bibitem [{\citenamefont {Blok}\ \emph {et~al.}(2021)\citenamefont {Blok},
  \citenamefont {Ramasesh}, \citenamefont {Schuster}, \citenamefont
  {O’Brien}, \citenamefont {Kreikebaum}, \citenamefont {Dahlen},
  \citenamefont {Morvan}, \citenamefont {Yoshida}, \citenamefont {Yao},\ and\
  \citenamefont {Siddiqi}}]{blok2021}%
  \BibitemOpen
  \bibfield  {author} {\bibinfo {author} {\bibfnamefont {M.}~\bibnamefont
  {Blok}}, \bibinfo {author} {\bibfnamefont {V.}~\bibnamefont {Ramasesh}},
  \bibinfo {author} {\bibfnamefont {T.}~\bibnamefont {Schuster}}, \bibinfo
  {author} {\bibfnamefont {K.}~\bibnamefont {O’Brien}}, \bibinfo {author}
  {\bibfnamefont {J.}~\bibnamefont {Kreikebaum}}, \bibinfo {author}
  {\bibfnamefont {D.}~\bibnamefont {Dahlen}}, \bibinfo {author} {\bibfnamefont
  {A.}~\bibnamefont {Morvan}}, \bibinfo {author} {\bibfnamefont
  {B.}~\bibnamefont {Yoshida}}, \bibinfo {author} {\bibfnamefont
  {N.}~\bibnamefont {Yao}},\ and\ \bibinfo {author} {\bibfnamefont
  {I.}~\bibnamefont {Siddiqi}},\ }\href
  {http://dx.doi.org/10.1103/PhysRevX.11.021010} {\bibfield  {journal}
  {\bibinfo  {journal} {Phys. Rev. X}\ }\textbf {\bibinfo {volume} {11}}
  (\bibinfo {year} {2021})}\BibitemShut {NoStop}%
\bibitem [{\citenamefont {Morvan}\ \emph {et~al.}(2021)\citenamefont {Morvan},
  \citenamefont {Ramasesh}, \citenamefont {Blok}, \citenamefont {Kreikebaum},
  \citenamefont {O'Brien}, \citenamefont {Chen}, \citenamefont {Mitchell},
  \citenamefont {Naik}, \citenamefont {Santiago},\ and\ \citenamefont
  {Siddiqi}}]{morvan2021}%
  \BibitemOpen
  \bibfield  {author} {\bibinfo {author} {\bibfnamefont {A.}~\bibnamefont
  {Morvan}}, \bibinfo {author} {\bibfnamefont {V.~V.}\ \bibnamefont
  {Ramasesh}}, \bibinfo {author} {\bibfnamefont {M.~S.}\ \bibnamefont {Blok}},
  \bibinfo {author} {\bibfnamefont {J.~M.}\ \bibnamefont {Kreikebaum}},
  \bibinfo {author} {\bibfnamefont {K.}~\bibnamefont {O'Brien}}, \bibinfo
  {author} {\bibfnamefont {L.}~\bibnamefont {Chen}}, \bibinfo {author}
  {\bibfnamefont {B.~K.}\ \bibnamefont {Mitchell}}, \bibinfo {author}
  {\bibfnamefont {R.~K.}\ \bibnamefont {Naik}}, \bibinfo {author}
  {\bibfnamefont {D.~I.}\ \bibnamefont {Santiago}},\ and\ \bibinfo {author}
  {\bibfnamefont {I.}~\bibnamefont {Siddiqi}},\ }\href
  {https://doi.org/10.1103/PhysRevLett.126.210504} {\bibfield  {journal}
  {\bibinfo  {journal} {Phys. Rev. Lett.}\ }\textbf {\bibinfo {volume} {126}},\
  \bibinfo {pages} {210504} (\bibinfo {year} {2021})}\BibitemShut {NoStop}%
\bibitem [{\citenamefont {Rigetti}\ and\ \citenamefont
  {Devoret}(2010)}]{rigetti2010}%
  \BibitemOpen
  \bibfield  {author} {\bibinfo {author} {\bibfnamefont {C.}~\bibnamefont
  {Rigetti}}\ and\ \bibinfo {author} {\bibfnamefont {M.}~\bibnamefont
  {Devoret}},\ }\href {https://doi.org/10.1103/PhysRevB.81.134507} {\bibfield
  {journal} {\bibinfo  {journal} {Phys. Rev. B}\ }\textbf {\bibinfo {volume}
  {81}},\ \bibinfo {pages} {134507} (\bibinfo {year} {2010})}\BibitemShut
  {NoStop}%
\bibitem [{\citenamefont {Kreikebaum}\ \emph {et~al.}(2020)\citenamefont
  {Kreikebaum}, \citenamefont {O’Brien}, \citenamefont {Morvan},\ and\
  \citenamefont {Siddiqi}}]{kreikebaum2020improving}%
  \BibitemOpen
  \bibfield  {author} {\bibinfo {author} {\bibfnamefont {J.}~\bibnamefont
  {Kreikebaum}}, \bibinfo {author} {\bibfnamefont {K.}~\bibnamefont
  {O’Brien}}, \bibinfo {author} {\bibfnamefont {A.}~\bibnamefont {Morvan}},\
  and\ \bibinfo {author} {\bibfnamefont {I.}~\bibnamefont {Siddiqi}},\ }\href
  {https://doi.org/10.1088/1361-6668/ab8617} {\bibfield  {journal} {\bibinfo
  {journal} {Supercond. Sci. Technol.}\ }\textbf {\bibinfo {volume} {33}},\
  \bibinfo {pages} {06LT02} (\bibinfo {year} {2020})}\BibitemShut {NoStop}%
\bibitem [{\citenamefont {Hertzberg}\ \emph {et~al.}(2021)\citenamefont
  {Hertzberg}, \citenamefont {Zhang}, \citenamefont {Rosenblatt}, \citenamefont
  {Magesan}, \citenamefont {Smolin}, \citenamefont {Yau}, \citenamefont
  {Adiga}, \citenamefont {Sandberg}, \citenamefont {Brink}, \citenamefont
  {Chow} \emph {et~al.}}]{hertzberg2020laserannealing}%
  \BibitemOpen
  \bibfield  {author} {\bibinfo {author} {\bibfnamefont {J.~B.}\ \bibnamefont
  {Hertzberg}}, \bibinfo {author} {\bibfnamefont {E.~J.}\ \bibnamefont
  {Zhang}}, \bibinfo {author} {\bibfnamefont {S.}~\bibnamefont {Rosenblatt}},
  \bibinfo {author} {\bibfnamefont {E.}~\bibnamefont {Magesan}}, \bibinfo
  {author} {\bibfnamefont {J.~A.}\ \bibnamefont {Smolin}}, \bibinfo {author}
  {\bibfnamefont {J.-B.}\ \bibnamefont {Yau}}, \bibinfo {author} {\bibfnamefont
  {V.~P.}\ \bibnamefont {Adiga}}, \bibinfo {author} {\bibfnamefont
  {M.}~\bibnamefont {Sandberg}}, \bibinfo {author} {\bibfnamefont
  {M.}~\bibnamefont {Brink}}, \bibinfo {author} {\bibfnamefont {J.~M.}\
  \bibnamefont {Chow}}, \emph {et~al.},\ }\href
  {https://doi.org/10.1038/s41534-021-00464-5} {\bibfield  {journal} {\bibinfo
  {journal} {npj Quantum Information}\ }\textbf {\bibinfo {volume} {7}},\
  \bibinfo {pages} {1} (\bibinfo {year} {2021})}\BibitemShut {NoStop}%
\bibitem [{\citenamefont {Fowler}\ \emph {et~al.}(2012)\citenamefont {Fowler},
  \citenamefont {Mariantoni}, \citenamefont {Martinis},\ and\ \citenamefont
  {Cleland}}]{fowler2012}%
  \BibitemOpen
  \bibfield  {author} {\bibinfo {author} {\bibfnamefont {A.~G.}\ \bibnamefont
  {Fowler}}, \bibinfo {author} {\bibfnamefont {M.}~\bibnamefont {Mariantoni}},
  \bibinfo {author} {\bibfnamefont {J.~M.}\ \bibnamefont {Martinis}},\ and\
  \bibinfo {author} {\bibfnamefont {A.~N.}\ \bibnamefont {Cleland}},\ }\href
  {http://dx.doi.org/10.1103/PhysRevA.86.032324} {\bibfield  {journal}
  {\bibinfo  {journal} {Physical Review A}\ }\textbf {\bibinfo {volume} {86}}
  (\bibinfo {year} {2012})}\BibitemShut {NoStop}%
\bibitem [{\citenamefont {Mitchell}\ \emph {et~al.}(2021)\citenamefont
  {Mitchell}, \citenamefont {Naik}, \citenamefont {Morvan}, \citenamefont
  {Hashim}, \citenamefont {Kreikebaum}, \citenamefont {Marinelli},
  \citenamefont {Lavrijsen}, \citenamefont {Nowrouzi}, \citenamefont
  {Santiago},\ and\ \citenamefont {Siddiqi}}]{mitchell2021hardwareefficient}%
  \BibitemOpen
  \bibfield  {author} {\bibinfo {author} {\bibfnamefont {B.~K.}\ \bibnamefont
  {Mitchell}}, \bibinfo {author} {\bibfnamefont {R.~K.}\ \bibnamefont {Naik}},
  \bibinfo {author} {\bibfnamefont {A.}~\bibnamefont {Morvan}}, \bibinfo
  {author} {\bibfnamefont {A.}~\bibnamefont {Hashim}}, \bibinfo {author}
  {\bibfnamefont {J.~M.}\ \bibnamefont {Kreikebaum}}, \bibinfo {author}
  {\bibfnamefont {B.}~\bibnamefont {Marinelli}}, \bibinfo {author}
  {\bibfnamefont {W.}~\bibnamefont {Lavrijsen}}, \bibinfo {author}
  {\bibfnamefont {K.}~\bibnamefont {Nowrouzi}}, \bibinfo {author}
  {\bibfnamefont {D.~I.}\ \bibnamefont {Santiago}},\ and\ \bibinfo {author}
  {\bibfnamefont {I.}~\bibnamefont {Siddiqi}},\ }\href
  {https://doi.org/10.1103/PhysRevLett.127.200502} {\bibfield  {journal}
  {\bibinfo  {journal} {Phys. Rev. Lett.}\ }\textbf {\bibinfo {volume} {127}},\
  \bibinfo {pages} {200502} (\bibinfo {year} {2021})}\BibitemShut {NoStop}%
\bibitem [{\citenamefont {Magesan}\ and\ \citenamefont
  {Gambetta}(2020)}]{magesan2020}%
  \BibitemOpen
  \bibfield  {author} {\bibinfo {author} {\bibfnamefont {E.}~\bibnamefont
  {Magesan}}\ and\ \bibinfo {author} {\bibfnamefont {J.~M.}\ \bibnamefont
  {Gambetta}},\ }\href {https://doi.org/10.1103/PhysRevA.101.052308} {\bibfield
   {journal} {\bibinfo  {journal} {Phys. Rev. A}\ }\textbf {\bibinfo {volume}
  {101}},\ \bibinfo {pages} {052308} (\bibinfo {year} {2020})}\BibitemShut
  {NoStop}%
\bibitem [{\citenamefont {Tripathi}\ \emph {et~al.}(2019)\citenamefont
  {Tripathi}, \citenamefont {Khezri},\ and\ \citenamefont
  {Korotkov}}]{tripathi2019}%
  \BibitemOpen
  \bibfield  {author} {\bibinfo {author} {\bibfnamefont {V.}~\bibnamefont
  {Tripathi}}, \bibinfo {author} {\bibfnamefont {M.}~\bibnamefont {Khezri}},\
  and\ \bibinfo {author} {\bibfnamefont {A.~N.}\ \bibnamefont {Korotkov}},\
  }\href {http://dx.doi.org/10.1103/PhysRevA.100.012301} {\bibfield  {journal}
  {\bibinfo  {journal} {Physical Review A}\ }\textbf {\bibinfo {volume} {100}}
  (\bibinfo {year} {2019})}\BibitemShut {NoStop}%
\bibitem [{\citenamefont {Sundaresan}\ \emph {et~al.}(2020)\citenamefont
  {Sundaresan}, \citenamefont {Lauer}, \citenamefont {Pritchett}, \citenamefont
  {Magesan}, \citenamefont {Jurcevic},\ and\ \citenamefont
  {Gambetta}}]{sundaresan2020}%
  \BibitemOpen
  \bibfield  {author} {\bibinfo {author} {\bibfnamefont {N.}~\bibnamefont
  {Sundaresan}}, \bibinfo {author} {\bibfnamefont {I.}~\bibnamefont {Lauer}},
  \bibinfo {author} {\bibfnamefont {E.}~\bibnamefont {Pritchett}}, \bibinfo
  {author} {\bibfnamefont {E.}~\bibnamefont {Magesan}}, \bibinfo {author}
  {\bibfnamefont {P.}~\bibnamefont {Jurcevic}},\ and\ \bibinfo {author}
  {\bibfnamefont {J.~M.}\ \bibnamefont {Gambetta}},\ }\href
  {http://dx.doi.org/10.1103/PRXQuantum.1.020318} {\bibfield  {journal}
  {\bibinfo  {journal} {PRX Quantum}\ }\textbf {\bibinfo {volume} {1}}
  (\bibinfo {year} {2020})}\BibitemShut {NoStop}%
\bibitem [{\citenamefont {Xiong}\ \emph {et~al.}(2021)\citenamefont {Xiong},
  \citenamefont {Ficheux}, \citenamefont {Somoroff}, \citenamefont {Nguyen},
  \citenamefont {Dogan}, \citenamefont {Rosenstock}, \citenamefont {Wang},
  \citenamefont {Nesterov}, \citenamefont {Vavilov},\ and\ \citenamefont
  {Manucharyan}}]{xiong2021}%
  \BibitemOpen
  \bibfield  {author} {\bibinfo {author} {\bibfnamefont {H.}~\bibnamefont
  {Xiong}}, \bibinfo {author} {\bibfnamefont {Q.}~\bibnamefont {Ficheux}},
  \bibinfo {author} {\bibfnamefont {A.}~\bibnamefont {Somoroff}}, \bibinfo
  {author} {\bibfnamefont {L.~B.}\ \bibnamefont {Nguyen}}, \bibinfo {author}
  {\bibfnamefont {E.}~\bibnamefont {Dogan}}, \bibinfo {author} {\bibfnamefont
  {D.}~\bibnamefont {Rosenstock}}, \bibinfo {author} {\bibfnamefont
  {C.}~\bibnamefont {Wang}}, \bibinfo {author} {\bibfnamefont {K.~N.}\
  \bibnamefont {Nesterov}}, \bibinfo {author} {\bibfnamefont {M.~G.}\
  \bibnamefont {Vavilov}},\ and\ \bibinfo {author} {\bibfnamefont {V.~E.}\
  \bibnamefont {Manucharyan}},\ }\href@noop {} {\bibinfo {title} {Arbitrary
  controlled-phase gate on fluxonium qubits using differential {AC-Stark}
  shifts}} (\bibinfo {year} {2021}),\ \Eprint
  {https://arxiv.org/abs/2103.04491} {arXiv:2103.04491 [quant-ph]} \BibitemShut
  {NoStop}%
\bibitem [{\citenamefont {Wei}\ \emph {et~al.}(2021)\citenamefont {Wei},
  \citenamefont {Magesan}, \citenamefont {Lauer}, \citenamefont {Srinivasan},
  \citenamefont {Bogorin}, \citenamefont {Carnevale}, \citenamefont {Keefe},
  \citenamefont {Kim}, \citenamefont {Klaus}, \citenamefont {Landers},
  \citenamefont {Sundaresan}, \citenamefont {Wang}, \citenamefont {Zhang},
  \citenamefont {Steffen}, \citenamefont {Dial}, \citenamefont {McKay},\ and\
  \citenamefont {Kandala}}]{wei2021quantum}%
  \BibitemOpen
  \bibfield  {author} {\bibinfo {author} {\bibfnamefont {K.~X.}\ \bibnamefont
  {Wei}}, \bibinfo {author} {\bibfnamefont {E.}~\bibnamefont {Magesan}},
  \bibinfo {author} {\bibfnamefont {I.}~\bibnamefont {Lauer}}, \bibinfo
  {author} {\bibfnamefont {S.}~\bibnamefont {Srinivasan}}, \bibinfo {author}
  {\bibfnamefont {D.~F.}\ \bibnamefont {Bogorin}}, \bibinfo {author}
  {\bibfnamefont {S.}~\bibnamefont {Carnevale}}, \bibinfo {author}
  {\bibfnamefont {G.~A.}\ \bibnamefont {Keefe}}, \bibinfo {author}
  {\bibfnamefont {Y.}~\bibnamefont {Kim}}, \bibinfo {author} {\bibfnamefont
  {D.}~\bibnamefont {Klaus}}, \bibinfo {author} {\bibfnamefont
  {W.}~\bibnamefont {Landers}}, \bibinfo {author} {\bibfnamefont
  {N.}~\bibnamefont {Sundaresan}}, \bibinfo {author} {\bibfnamefont
  {C.}~\bibnamefont {Wang}}, \bibinfo {author} {\bibfnamefont {E.~J.}\
  \bibnamefont {Zhang}}, \bibinfo {author} {\bibfnamefont {M.}~\bibnamefont
  {Steffen}}, \bibinfo {author} {\bibfnamefont {O.~E.}\ \bibnamefont {Dial}},
  \bibinfo {author} {\bibfnamefont {D.~C.}\ \bibnamefont {McKay}},\ and\
  \bibinfo {author} {\bibfnamefont {A.}~\bibnamefont {Kandala}},\ }\href@noop
  {} {\bibinfo {title} {Quantum crosstalk cancellation for fast entangling
  gates and improved multi-qubit performance}} (\bibinfo {year} {2021}),\
  \Eprint {https://arxiv.org/abs/2106.00675} {arXiv:2106.00675 [quant-ph]}
  \BibitemShut {NoStop}%
\bibitem [{\citenamefont {{Gurobi Optimization, LLC}}(2021)}]{gurobi}%
  \BibitemOpen
  \bibfield  {author} {\bibinfo {author} {\bibnamefont {{Gurobi Optimization,
  LLC}}},\ }\href {https://www.gurobi.com} {\bibinfo {title} {{Gurobi Optimizer
  Reference Manual}}} (\bibinfo {year} {2021})\BibitemShut {NoStop}%
\bibitem [{\citenamefont {Hart}\ \emph {et~al.}(2011)\citenamefont {Hart},
  \citenamefont {Watson},\ and\ \citenamefont {Woodruff}}]{pyomo}%
  \BibitemOpen
  \bibfield  {author} {\bibinfo {author} {\bibfnamefont {W.~E.}\ \bibnamefont
  {Hart}}, \bibinfo {author} {\bibfnamefont {J.-P.}\ \bibnamefont {Watson}},\
  and\ \bibinfo {author} {\bibfnamefont {D.~L.}\ \bibnamefont {Woodruff}},\
  }\href {https://doi.org/10.1007/s12532-011-0026-8} {\bibfield  {journal}
  {\bibinfo  {journal} {Math. Program. Comput.}\ }\textbf {\bibinfo {volume}
  {3}},\ \bibinfo {pages} {219} (\bibinfo {year} {2011})}\BibitemShut {NoStop}%
\bibitem [{\citenamefont {Chamberland}\ \emph {et~al.}(2020)\citenamefont
  {Chamberland}, \citenamefont {Zhu}, \citenamefont {Yoder}, \citenamefont
  {Hertzberg},\ and\ \citenamefont {Cross}}]{chamberland2020}%
  \BibitemOpen
  \bibfield  {author} {\bibinfo {author} {\bibfnamefont {C.}~\bibnamefont
  {Chamberland}}, \bibinfo {author} {\bibfnamefont {G.}~\bibnamefont {Zhu}},
  \bibinfo {author} {\bibfnamefont {T.~J.}\ \bibnamefont {Yoder}}, \bibinfo
  {author} {\bibfnamefont {J.~B.}\ \bibnamefont {Hertzberg}},\ and\ \bibinfo
  {author} {\bibfnamefont {A.~W.}\ \bibnamefont {Cross}},\ }\href
  {https://doi.org/10.1103/PhysRevX.10.011022} {\bibfield  {journal} {\bibinfo
  {journal} {Phys. Rev. X}\ }\textbf {\bibinfo {volume} {10}},\ \bibinfo
  {pages} {011022} (\bibinfo {year} {2020})}\BibitemShut {NoStop}%
\end{thebibliography}%

%%%%%%%%%%%%%%%%%%%%%%%%%%%%%%%%%%%%%%%%%%%%%%%%%%%%%%%%%%%%%%%%%%%%%%%%
\vfill
\framebox{\parbox{.90\linewidth}{\scriptsize The submitted manuscript has been created by
UChicago Argonne, LLC, Operator of Argonne National Laboratory (``Argonne'').
Argonne, a U.S.\ Department of Energy Office of Science laboratory, is operated
under Contract No.\ DE-AC02-06CH11357.  The U.S.\ Government retains for itself,
and others acting on its behalf, a paid-up nonexclusive, irrevocable worldwide
license in said article to reproduce, prepare derivative works, distribute
copies to the public, and perform publicly and display publicly, by or on
behalf of the Government.  The Department of Energy will provide public access
to these results of federally sponsored research in accordance with the DOE
Public Access Plan \url{http://energy.gov/downloads/doe-public-access-plan}.}}

\clearpage
%%%%%%%%%% Merge with supplemental materials %%%%%%%%%%
%%%%%%%%%% Prefix a "S" to all equations, figures, tables and reset the counter %%%%%%%%%%
\setcounter{equation}{0}
\setcounter{figure}{0}
\setcounter{table}{0}
\setcounter{page}{1}
\makeatletter
\renewcommand{\theequation}{S\arabic{equation}}
\renewcommand{\thetable}{S\arabic{table}}
\renewcommand{\thefigure}{S\arabic{figure}}
\renewcommand{\bibnumfmt}[1]{[S#1]}
\renewcommand{\citenumfont}[1]{S#1}
%%%%%%%%%% Prefix a "S" to all equations, figures, tables and reset the counter %%%%%%%%%%

% Appendix
\section{Transmon model}
    Transmons are multilevel systems that can be approximated by a weakly
anharmonic oscillator. The energy of the $n$th level is given by
\begin{equation}
	E_n/h =  f_{n} = n f_0 + n(n-1)\frac{\alpha}{2},
\end{equation}
where $h$ is the Planck constant, $f_0$ is the frequency of the $0\mapsto 1$
transition, and $\alpha$ is the anharmonicity. In practice, $f_0$ is within
$4.5--6$ GHz, and $\alpha$ is typically between $-200$ and $-350$ MHz.
The transition frequencies are then given by
\begin{equation}
    f_{ij} = f_i - f_j = (i-j)f_0 + (i-j)(i+j-1)\frac{\alpha}{2}.
\end{equation}
These transition frequencies are considered when looking at
collisions. The transition to avoid is the transition that goes from a state
that can be populated (0 and 1 for qubits) to another state. In some situations,
2-photon transitions also have to be considered; in this case, the transition
frequency is given by half the frequency of the 2-photon transition.

\section{Constraints for qubit architecture}
    We now describe the frequency constraints and collisions on a
CR architecture where the transmons are used as qubits. The
constraints are listed in Table I of the main text.

%\onecolumngrid
\begin{figure}
  \begin{center}
        \includegraphics{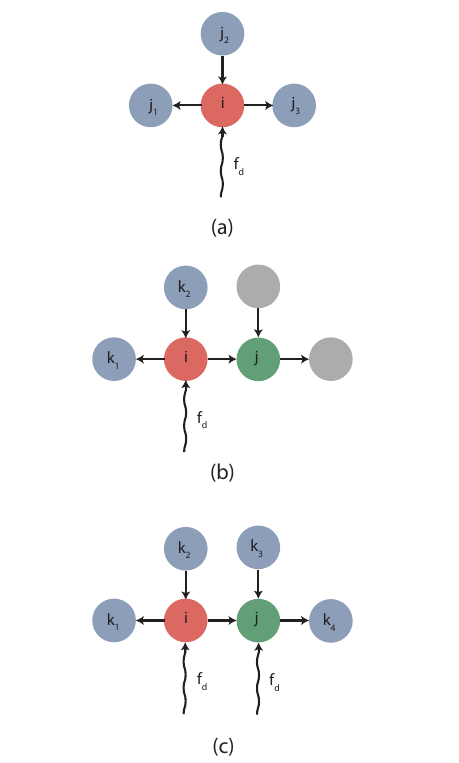}. % todo, maybe remove the with?
    \end{center}
    \caption{
    Configuration of the different sites considered. (a) Addressability constraint.    The transmon $i$ is neighbors to $j_1$,$j_2$,$j_2$. When applying a pulse at the $\ket{0} \rightarrow \ket{1}$ 
    or $\ket{1} \rightarrow \ket{2}$ transition of $i$, the frequency of the pulse must avoid the transition frequencies 
    of these neighbors. (b) Spectators constraints for the CR architecture. (c) Spectator constraints for the CZ architecture.}
\end{figure}  
% \twocolumngrid

\subsection{Spectators of the single-qubit gates}
Manipulating a qubit, that is sending a pulse at its transition between
the $0$ and the 1 state, should not drive its neighbors' transmons transitions.
Since qubits populate only the states 0 and 1, we have to avoid the 0-1 transition
and the 1 to 2 transitions. This leads to two constraints that we will refer to as
A constraints:
\begin{itemize}
  \item Type A1: $f_d=f_i$. Avoid the $0 \mapsto 1$ transitions of $j$:
    \begin{equation}
      \left| f_d - f_j \right| \ge \delta_{A1} \qquad \forall (i,j) \in E
    \end{equation}
  \item Type A2: $f_d=f_i$. Avoid the $1 \mapsto 2$ transitions of $j$:
    \begin{equation}
      \left| f_d - f_j - \alpha_j \right| \ge \delta_{A2} \qquad \forall (i,j) \in E
    \end{equation}
\end{itemize}
To code these constraints, we notice that the constraint A1 can be applied to
the directed graph $\vec{E}$, whereas because of the asymmetry in A2, we have cut
the second constraint into two constraints: one as described in the equation
and another with the inversion of the two indices $i$ and $j$ to be applied to the oriented edges.

\subsection{Entanglement with CR constraints}

We have a constraint on the frequency difference for an interaction gate to
be fast enough:
\begin{itemize}
    \item Type C1: Gate fast enough for $f_d=f_j$:
        \begin{equation}
                  f_{i} + \alpha_i \le f_d  \le f_i  \qquad \forall (i,j) \in \vec{E}.
        \end{equation}
\end{itemize}
For our optimization scheme, we introduce $\delta_{C1} = 5$ MHz to control this constraint in the objective function:
\begin{equation}
    f_d - f_{i} - \alpha_i  \ge \delta_{C1}  \qquad \forall (i,j) \in \vec{E}
\end{equation}
and 
\begin{equation}
    f_{i}  - f_d \ge \delta_{C1}  \qquad \forall (i,j) \in \vec{E}.
\end{equation}

The entanglement gate is performed with the CR interaction where a
pulse at the frequency of the \textit{target} transmon $j$ is applied to the
\textit{control} transmon $i$. One requirement for this gate to work
efficiently is to not affect the control transmon, meaning that the frequency
of the applied pulse has to be far enough from the various transition of the
control transmon. These constraints apply to the oriented edges of the graph
$\vec{E}$. There is a collision when the drive frequency $f_d$ collides with a
transition of the control qubits. Since the drive is strong, we must also 
avoid the 2-photon transitions as well as the 1-photon transitions:

\begin{itemize}
  \item Type E1: $f_d$ must avoid the $0 \mapsto 1$ transition of $i$:
    \begin{equation}
      \left| f_d - f_i \right| \ge \delta_{E1} \qquad \forall e = (i,j) \in \vec{E}
    \end{equation}
  \item Type E2: $f_d$ must avoid the $1 \mapsto 2$ transition of $i$:
    \begin{equation}
      \left| f_d - f_i - \alpha_i \right| \ge \delta_{E2} \qquad \forall (i,j) \in \vec{E}
    \end{equation}
  \item Type D1: $f_d$ must avoid the $0 \mapsto 2$ 2-photon transitions of $i$:
    \begin{equation}
      \left| f_d - f_i - \alpha_i/2\right| \ge \delta_{D1} \qquad \forall (i,j) \in \vec{E}
    \end{equation}
\end{itemize}

We note that since $f_d = f_j$, the constraints E1 and E2 are included in
the A1 and A2 constraints. We include them because they decrease
the fidelity of different gates (single-qubit gate and entanglement gate) and
the distinction will be even more relevant for the off-resonant driving used
for the SiZZle gate.

\subsection{Entanglement pulses: spectators}

Another constraint when performing the entanglement drive is to not influence the
other neighbors' transmons. This is related to the addressability, but in the
context of entanglement gates, and is usually referred to as a spectator collision.
The conditions are then identical to the addressability constraints $A_n$ but
now with the frequency drive of the
entanglement gate $f_d$, and the set of transmons affected is given by the set
of transmons neighbors to the transmon where the drive is applied. We denote
this set by $N$. For the CR gate, the pulse is applied only to the
control transmon $i$ of the edge $(i, j)$, so it includes only the neighbors of
$i$ that are not $j$.
\begin{equation}
\begin{aligned}
    \cN = \{&(i, j, k) \in V^3:  (i,j)\in \vec{E}  \text{ with } (i,k) \in \vec{E} \text{ or } \\ &(k,i) \in \vec{E} \text{ and } j \neq k\}
\end{aligned}
\end{equation}

\begin{itemize}
  \item Type S1: $f_d$ must avoid the $0 \mapsto 1$ transitions of $k$:
    \begin{equation}
      \left| f_d - f_k \right| \ge \delta_{S1} \qquad \forall (i,j, k) \in \cN
    \end{equation}
  \item Type S2: $f_d$ must avoid the $1 \mapsto 2$ transitions of $k$:
    \begin{equation}
      \left| f_d - f_k - \alpha_k \right| \ge \delta_{S2} \qquad \forall (i,j, k) \in \cN
    \end{equation}
\end{itemize}

A more subtle type of collision can occur when the sum of the frequencies of
the target and a neighbor is equal to the frequency of the $\ket{0}
\rightarrow \ket{2}$ 2-photon transition. This is expressed with
\begin{itemize}
  \item Type T1: The sum of neighbor frequencies must avoid 
  the $\ket{0} \rightarrow \ket{2}$ transition of $k$:
  
  \begin{equation}
    \begin{aligned}
      \left| f_d + f_{k} - 2f_{i}-\alpha_i  \right| \ge \delta_{T1} \qquad \forall (i,j, k) \in \cN.
    \end{aligned}
    \end{equation}
\end{itemize}

    % \input{sections/supp02.tex}

% Small system analysis
% \section{Constraints for qutrit architecture}
    % \input{sections/supp03.tex}

\section{Constraints for CR qutrit architecture}
    Following \cite{blok2021, morvan2021}, a qutrit processor can be realized on
fixed-frequency transmon architectures. Universal single-qutrit control can be
realized through the use of the $\ket{0} \mapsto \ket{1}$ and the $\ket{1}
\mapsto \ket{2}$ transitions. The entanglement gate can be realized with the
cross-resonance interaction with these same transitions of the target
transmons. Therefore, mobilizing qutrits requires adding these new drives to
the constraint list. We list them in \Cref{tab:constraints_qutrit}. For
the addressability constraints, we split the constraints into two categories:
the  $\ket{0} \mapsto \ket{1}$ drive leads to the usual $A_i$ constraints, and
the $\ket{1} \mapsto \ket{2}$ drive leads to the $B_i$ constraints. Note that
because the state $\ket{2}$ can now be populated, the unwanted driving of the
transition $\ket{2} \mapsto {3}$ transition needs to be considered. For the
entanglement and spectator constraints, we have simply parametrized the drive
frequency in the constraints and both drive frequencies need to be considered.
We note that one possible architecture (less constrained than a full-qutrit
processor) is to use the $\ket{1} \mapsto \ket{2}$ transition or
single-transmon drive and use only the $f_d = f_j$ entanglement drive. This
type of architecture can also be useful and removes several constraints but
requires more overhead.

\begin{table*}[htbp]
    \caption{\label{tab:constraints_qutrit} Constraints imposed by the
    architecture on the frequency allocation on the graph for a CR
    qutrit architecture, following the notation of Table I. The indices
    refer to the participants and are defined by the graph underlying the
    processors. $\vec{E}$ represents the oriented graph, and $E$ represents the
    unoriented graph, meaning that each edge is taken into account twice, once in
    each direction. The frequency of the drive microwave $f_d$ can take two
    values for the qutrit case: the frequency of the target qubit $f_d = f_j$
    and the frequency of the $\ket{1} \mapsto \ket{2}$ transition $f_d =
    f_j+\alpha_j$. Since there is no full analysis of the performance of the
    qutrit CR gate as a function of the threshold, we have used the same values
    for the bounds as for the qubit case.}
\begin{tabular}{|cc|cccc|c|}
\hline
               &    &  & Definition                                                                 & Participants        & Bounds & Drive Frequency        \\ \hline
Addressability & A1 &  & $\left| f_i - f_j \right| \ge \delta_{A1}$                                 & $(i,j) \in E$       & 17 MHz &                        \\
               & A2 &  & $\left| f_i - f_j - \alpha_j \right| \ge \delta_{A2}$                      & $(i,j) \in E$       & 30 MHz &                        \\
               & A3 &  & $\left| f_i - f_j - 2\alpha_j \right| \ge \delta_{A3}$                     & $(i,j) \in E$       & 30 MHz &                        \\
               & B1 &  & $\left| f_i +\alpha_i - f_j \right| \ge \delta_{B1}$                       & $(i,j) \in E$       & 30 MHz &                        \\
               & B2 &  & $\left| f_i+\alpha_i - f_j-\alpha_j \right| \ge \delta_{B2}$               & $(i,j) \in E$       & 17 MHz &                        \\
               & B3 &  & $\left| f_i+\alpha_i - f_j-2\alpha_j \right| \ge \delta_{B3}$              & $(i,j) \in E$       & 30 MHz &                        \\ \hline
Entanglement   & C1 &  & $f_{i} + \alpha_i \le f_d \le f_i $                                        & $(i,j) \in \vec{E}$ & ---    &                        \\
               & E1 &  & $\left| f_d - f_i \right| \ge \delta_{E1}$                                 & $(i,j) \in \vec{E}$ & 17 MHz &                        \\
               & E2 &  & $\left| f_d - f_i - \alpha_i \right| \ge \delta_{E2}$                      & $(i,j) \in \vec{E}$ & 30 MHz &                        \\
               & E3 &  & $\left| f_d - f_i - 2\alpha_i \right| \ge \delta_{E3}$                     & $(i,j) \in \vec{E}$ & 30 MHz &                        \\
               & D1 &  & $\left| f_d - f_i - \alpha_i/2\right| \ge \delta_{D1}$                     & $(i,j) \in \vec{E}$ & 2 MHz  & $f_d = f_j$            \\
               & D2 &  & $\left| f_d - f_i - 3\alpha_i/2\right| \ge \delta_{D2}$                    & $(i,j) \in \vec{E}$ & 2 MHz  & and                    \\ \cline{1-6}
Spectators     & S1 &  & $\left| f_d - f_k \right| \ge \delta_{S1}$                                 & $(i,j, k) \in N$    & 17 MHz & $f_d = f_j + \alpha_j$ \\
               & S2 &  & $\left| f_d - f_k - \alpha_k \right| \ge \delta_{S2}$                      & $(i,j, k) \in N$    & 25 MHz &                        \\
               & S3 &  & $\left| f_d - f_k - 2\alpha_k \right| \ge \delta_{S3}$                     & $(i,j, k) \in N$    & 25 MHz &                        \\
               & T1 &  & $\left| f_{d} + f_{k} - 2f_{i}-\alpha_i\right| \ge \delta_{T1} $           & $(i,j, k) \in N$    & 17 MHz &                        \\
               & T2 &  & $\left| f_{d} + f_{k}+\alpha_k - 2f_{i}-3\alpha_i\right| \ge \delta_{T2} $ & $(i,j, k) \in N$    & 17 MHz &                        \\ \hline
\end{tabular}
\end{table*}

\section{Constraints for CZ qubit architecture}
    The differential AC-Stark shift considerably changes the constraints of the
architecture: The addressability is unchanged compared with the qubit CR case,
but the drive frequency of the entanglement gate can now take any frequency and
is not set by the target transmon. This introduces an extra degree of freedom in
the problem. We now need to find out whether there exists a drive satisfying the
constraints for a given layout. This differential AC-Stark shift requires 
driving both qubits at the drive frequency. This leads to new constraints
for the entanglement drive involving the neighbors of the second qubit,
constraints that were absent from the one-qubit case. 
\Cref{tab:constraints_cz} lists the constraints we have considered.

\begin{table*}[]
    \caption{\label{tab:constraints_cz} Constraints imposed by the
    architecture on the frequency allocation on the graph for a differential
    AC-Stark shift qubit architecture, following the notation of Table I. The
    indices refer to the participants and are defined by the graph underlying the
    processors. $\vec{E}$ represents the oriented graph, and $E$ represents the
    unoriented graph, meaning that each edge is taken into account twice in
    each direction. The frequency of the drive microwave $f_d$ can take any
    arbitrary value within the constraints. Since the AC-Stark shift is obtained
    by applying a drive on both qubits, we need to add the constraint with an additional 
    ``$t$'' that indicates that the target's frequency also needs to be part
    of the constraints. For the spectators' errors, we need to consider the
    neighbors of both qubits, depicted by the ensemble $\tilde{N}$ that
    includes the neighbors of both $i$ and $j$. Since there is no full analysis of
    the performance of the AC-Stark shift gate as a function of the
    thresholds, we have used the same values for the bounds as for the CR qubit
    case.}
\begin{tabular}{|cc|cccc|c|}
\hline
               &     &  & Definition                                                       & Participants             & Bounds & Drive Frequency            \\ \hline
Addressability & A1  &  & $\left| f_i - f_j \right| \ge \delta_{A1}$                       & $(i,j) \in E$            & 17 MHz &                            \\
               & A2  &  & $\left| f_i - f_j - \alpha_j \right| \ge \delta_{A2}$            & $(i,j) \in E$            & 30 MHz &                            \\ \hline
Entanglement   & C1  &  & $f_{i} + \alpha_i \le f_d \le f_i $                              & $(i,j) \in \vec{E}$      & ---    &                            \\
               & C1t &  & $f_{j} + \alpha_j \le f_d \le f_j $                              & $(i,j) \in \vec{E}$      & ---    &                            \\
               & E1  &  & $\left| f_d - f_i \right| \ge \delta_{E1}$                       & $(i,j) \in \vec{E}$      & 17 MHz &                            \\
               & E2  &  & $\left| f_d - f_i - \alpha_i \right| \ge \delta_{E2}$            & $(i,j) \in \vec{E}$      & 30 MHz &                            \\
               & E1t &  & $\left| f_d - f_j \right| \ge \delta_{E1}$                       & $(i,j) \in \vec{E}$      & 17 MHz & $f_d$  is a free parameter \\
               & E2t &  & $\left| f_d - f_j - \alpha_j \right| \ge \delta_{E2}$            & $(i,j) \in \vec{E}$      & 30 MHz &                            \\
               & D1  &  & $\left| f_d - f_i - \alpha_i/2\right| \ge \delta_{D1}$           & $(i,j) \in \vec{E}$      & 2 MHz  &                            \\
               & D1t &  & $\left| f_d - f_j - \alpha_j/2\right| \ge \delta_{D1}$           & $(i,j) \in \vec{E}$      & 2 MHz  &                            \\ \cline{1-6}
Spectators     & S1  &  & $\left| f_d - f_k \right| \ge \delta_{S1}$                       & $(i,j, k) \in \tilde{N}$ & 17 MHz &                            \\
               & S2  &  & $\left| f_d - f_k - \alpha_k \right| \ge \delta_{S2}$            & $(i,j, k) \in \tilde{N}$ & 25 MHz &                            \\
               & T1  &  & $\left| f_{d} + f_{k} - 2f_{i}-\alpha_i\right| \ge \delta_{T1} $ & $(i,j, k) \in \tilde{N}$ & 17 MHz &                            \\ \hline
\end{tabular}
\end{table*}

\begin{figure*}
    \begin{center}
        \includegraphics{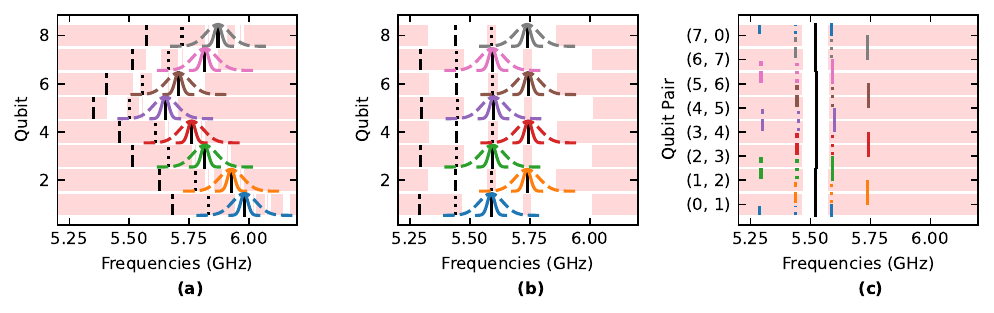}
    \end{center}
    \caption{\label{fig:ring_yield} (a) Collision regions of an 8-qubit ring for the qubit CR architecture.
    The solid black line indicates the $\ket{0} \rightarrow \ket{1}$, the dashed
    black line the $\ket{1} \rightarrow \ket{2}$ transition, and the dotted line
    the 2-photon transitions $\ket{0} \rightarrow \ket{2}$. The red regions
    indicate regions considered as collisions for the device. We have plotted
    the normal distribution of the optimized frequencies for dispersions of $\sigma_f = 15 (50)$ MHz with the dashed (solid) lines.
    (b) Corresponding collision regions of an 8-qubit for the qubit CZ architecture.
    (c) Allowed regions for the CZ drive applied to each qubit in a specific qubit pair. Since the drive is no longer restricted to be exactly equal to the target qubit frequency, as it is in the CR architecture, it can be adjusted in situ within the allowed region (white) once the actual qubit frequencies (colored) are known.}
\end{figure*}

\section{Optimization strategy}
    Finding an appropriate objective was an iterative process. Initially, we used $\delta_i$ to quantify the 
degree to which each set of constraints (on the respective frequencies and
anharmonicities) is satisfied.
Our first attempt was to maximize the sum of all $\delta_i$. For simple
connectivity graphs, the MIP solution produced poor yields due to $\delta_i$
being zero for some $i$.

We next tried the objective
\begin{equation}
  \label{eq:obj1_b}
  \maximize \sum_i (\delta_i - \bar{\delta}_i)
\end{equation}
while constraining $\delta_i \ge \bar{\delta}_i$ where $\bar{\delta}_i$ are the
bounds declared in 
Tables~\ref{tab:constraints_qutrit} and
\ref{tab:constraints_cz}. Unfortunately, this objective also resulted in some
of the $\delta_i$ taking their lower bound value $\bar{\delta}_i$. 
We also considered weighting the terms in \eqref{eq:obj1_b} by counting how
often each set of constraints corresponding to each $\delta_i$ is violated in a
collision. While we tried ad-hoc increases for the weights applied to terms with more collisions, we did not
find a consistent scheme for adjusting weights.

Because different terms in \eqref{eq:obj1_b} were zero in some solutions, we
considered a second step where we introduce a new scalar variable $K$ and optimize the objective
\begin{equation}
  \label{eq:obj2_b}
  \maximize K 
\end{equation}
subject to the additional constraints $K \le (\delta_i - \bar{\delta}_i)$ for each
group of constraints $i$. This produced an optimal value $K^*$. We then re-solved \eqref{eq:obj1_b}
with the additional constraints $K^* \le (\delta_i - \bar{\delta}_i)$ for each group of constraints $i$. This helped
drive all the $\delta_i$ away from their lower bounds (if possible) and
then allowed for an additional step of optimization.

This greatly improved the yield, so we considered a final polishing step to 
provide further improvement. After the above process, we set $K_i^*$ to be
the difference between each $\delta_i$ and its lower bound. We then took this quantity
and introduced additional variables $\delta_{i,e}$ for each $e \in E$ and solved
\begin{equation}
  \label{eq:obj3_b}
  \maximize \sum_i \sum_{e \in E} \delta_{i,e}
\end{equation}
subject to the constraints that $\delta_{i,e} \ge K_i^*$ from the second step.
    
\section{Scaling}
    \begin{figure}
    \begin{center}
        \includegraphics{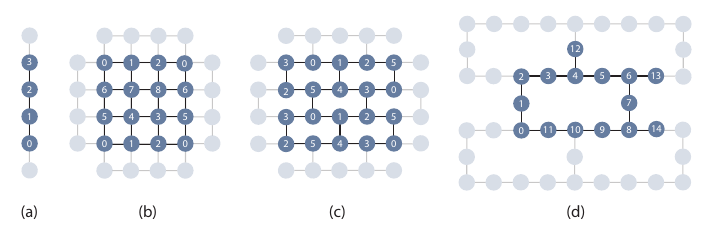}. % todo, maybe remove the with?
    \end{center}
    \caption{\label{fig:lattice_number} Site numbering for the lattice frequencies in Table \ref{tab:lattice-freq}. The unit cell for the linear chain (a) has size $n = 8$.}
\end{figure}    

\begin{table*}[]
    \caption{\label{tab:lattice-freq} Optimized frequencies for the lattices shown in Figure 2 in the main text. Qubit anharmonicities were fixed to 350 MHz for every site.}
\begin{tabular}{|c|c|c|c|}
\hline
Lattice & Site & CR Frequency & CZ Frequency \\ \hline
Linear  & 0  & 5.980 & 4.7385 \\
        & 1  & 5.925 & 4.5935 \\
        & 2  & 5.815 & 4.7435 \\
        & 3  & 5.760 & 4.5985 \\
        & 4  & 5.650 & 4.7435 \\
        & 5  & 5.705 & 4.5935 \\
        & 6  & 5.815 & 4.7385 \\
        & 7  & 5.870 & 4.5885 \\
        \hline
Square  & 0  & 5.8530 & 4.5748 \\
        & 1  & 5.7822 & 4.6296 \\
        & 2  & 5.9486 & 4.7522 \\
        & 3  & 5.7508 & 4.6296 \\
        & 4  & 5.7194 & 4.7392 \\
        & 5  & 5.9172 & 4.5748 \\
        & 6  & 5.8844 & 4.7522 \\
        & 7  & 5.8136 & 4.5748 \\
        & 8  & 5.9800 & 4.6296 \\
        \hline
Hexagon & 0  & 4.722 & 4.980 \\
        & 1  & 4.694 & 4.830 \\
        & 2  & 4.607 & 4.975 \\
        & 3  & 4.733 & 4.830 \\
        & 4  & 4.520 & 4.975 \\
        & 5  & 4.646 & 4.830 \\
        \hline
Heavy Hexagon
        & 0  & 4.575 & 4.731 \\
        & 1  & 4.630 & 4.596 \\
        & 2  & 4.520 & 4.731 \\
        & 3  & 4.630 & 4.596 \\
        & 4  & 4.575 & 4.731 \\
        & 5  & 4.685 & 4.596 \\
        & 6  & 4.630 & 4.731 \\
        & 7  & 4.685 & 4.596 \\
        & 8  & 4.575 & 4.731 \\
        & 9  & 4.630 & 4.596 \\
        & 10 & 4.520 & 4.731 \\
        & 11 & 4.630 & 4.596 \\
        & 12 & 4.630 & 4.731 \\
        & 13 & 4.685 & 4.596 \\
        & 14 & 4.630 & 4.596 \\
\hline
\end{tabular}
\end{table*}

\begin{figure}
    \begin{center}
        \includegraphics{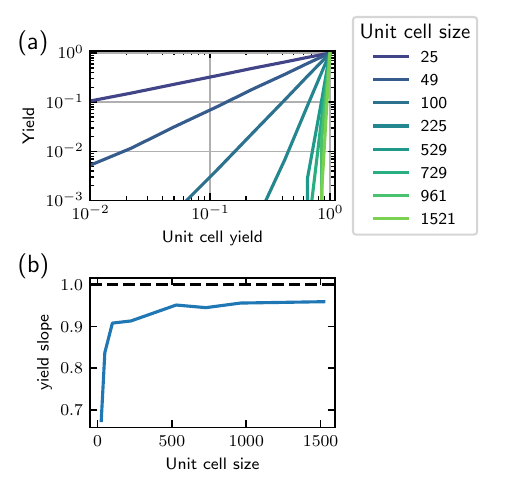}. % todo, maybe remove the with?
    \end{center}
    \caption{\label{fig:yield_scaling} Scaling properties of the yield as a
    function of the unit cells yield with a periodic boundary condition. (a) Shows that the
    yield of the large system scales as a power law of the cell yield. (b)
    Shows the scaling compared with the unit cell for the different sizes of
    the system. For the small system, border effects are strong; for the large
    system, the ratio is close to 1, indicating small border effects.}
\end{figure}

To check how the yield scales as we increase the lattice size, we calculate the yield for a system with periodic boundaries conditions. Each unit cell is a 5x5 square lattice of qubits and we use constraints for the CR architecture. We then tile this solution to create a larger $n \times n$ system and calculate the yield. Figure \ref{fig:yield_scaling} shows the evolution of the yield with an increasing number of sites. The linear correlation in the log-log plot shows that the yield for the large lattice can be calculated with a simple exponential law given in Eq. 2. In Figure \ref{fig:yield_scaling} (b) we show the correlation coefficient between the yield for the periodic unit cell and the yield for the larger system, normalized by the system size. For larger systems, the correlation is almost one; for smaller systems, the correlation is lower, which indicates that boundary effects play a significant role in the yield for smaller systems. As the system size increases, these boundary effects become less important.
We note this boundary effect is simply because some constraints don’t exist on the edges of the lattice, compared to the bulk. This reduces the number of constraints at the edges, and thus Eq. 2 of the main text gives an upper bound on the yield of the lattice.

\end{document}